\documentclass[a4paper,11pt]{article}
\usepackage{jheppub} 
\usepackage{lineno}
\usepackage{multirow}
\usepackage{bm}
\usepackage{tensor}
\usepackage{longtable}
\usepackage{xcolor}
 \usepackage[normalem]{ulem}

\toccontinuoustrue

\usepackage{amsmath}

\newcommand*{\defeq}{\stackrel{\text{def}}{=}}
\newcommand{\intwK}{\kappa^{-1}\! \!\int \!\! d \mathbf{x} \,}
\newcommand{\intnK}{\int \!\! d \mathbf{x} \,}

\newcommand{\p}{\partial}

\newcommand{\pc}{\partial_\frak{c}}
\newcommand{\poc}[1]{\partial_\frak{c} (#1)}
\newcommand{\ppc}[1]{\partial_\frak{c} #1}
\newcommand{\pp}{\partial_\frak{p}}
\newcommand{\pop}[1]{\partial_\frak{p} (#1)}
\newcommand{\ppp}[1]{\partial_\frak{p} #1}

\newcommand{\tr}[1]{\delta #1^b_b}

\newcommand{\iPB}[2]{$\left\{ \mathbb{#1}, \mathbb{#2} \right\}$}

\newcommand{\ddN}{\Delta \bigl[ \delta N \bigr]}

\title{\boldmath Anomaly freedom in effective Loop Quantum Cosmology refined: extended functional dependence of the counterterms}

\author{Maxime De Sousa, Aurélien Barrau, and Killian Martineau}
\affiliation{Laboratoire de Physique Subatomique et de Cosmologie, Universit\'e Grenoble-Alpes, CNRS/IN2P3\\
53, avenue des Martyrs, 38026 Grenoble cedex, France}

\abstract{Instead of assuming that they depend only on the background variables, we investigate the hypothesis that counterterms appearing in the deformed algebra approach to loop quantum cosmology depend on the full phase-space variables. We derive the associated anomalies and solve the entire system in several specific cases. New restrictions on the generalized holonomy corrections are obtained.}

\begin{document}
\maketitle
\allowdisplaybreaks
\section{Introduction}
Loop quantum gravity is a nonperturbative background-invariant tentative quantization of general relativity. Loop quantum cosmology (LQC) is its symmetry reduced version -- it actually refers to a number of different approaches -- applied to the decription of the Universe. In this article, we focus on the so-called deformed algebra approach of the holonomy-corrected effective theory and ask a simple question: how does a generalization of the functional dependence of the counterterms change the picture ?\\

Although counterterms were initially introduced to compensate for anomalies induced by holonomy corrections of the constraints, they can now be better viewed as terms widening the class of gravity theories considered, under the requirement that the connection appears only through its holonomy and that the algebra remains first class, that is consistent. Usually those counterterms are assumed to depend only on background variables. In this work, we explore the possibility that they do depend on the full phase-space variables.\\

To keep the article as brief as possible -- the calculations being quite involved and lengthy -- we refer the interested reader to our previous introductory and pedagogical work \cite{DeSousa:2024icf} (and to \cite{Han:2017wmt}) for all details, motivations, and a quite exhaustive bibliography. We deliberately keep, here, the focus on new insights only.
\section{Effective loop quantum cosmology}
    To keep this work self-contained, we quickly go through the basics (see \cite{Ashtekar:2011ni} for a review). Beyond the key background result, that is the replacement of the big bang singularity by a regular big bounce, important progresses was made in the investigation of cosmological perturbations.\\

In this article, we focus on the so-called deformed algebra approach \cite{Bojowald:2008jv, Wu:2012mh, Han:2018usc, BenAchour:2016leo, Mielczarek:2011ph, Cailleteau:2011kr, Cailleteau:2012fy, Cailleteau:2013kqa, Han:2017wmt, Bolliet:2015bka, Martineau:2017tdx, Barrau:2018gyz} with a clear focus on consistency issues. When the equations of gravity are corrected at the effective level it is hard to determine whether the subtle consistency conditions captured by the first-class nature of the algebra of constraints are still satisfied \cite{Barrau:2014maa}. Nice results were recently obtained for black holes \cite{BenAchour:2018khr, Bojowald:2018xxu, Bojowald:2020dkb, Arruga:2019kyd, Alonso-Bardaji:2023niu, Alonso-Bardaji:2022ear, Alonso-Bardaji:2020rxb, Alonso-Bardaji:2021tvy, Belfaqih:2024vfk} but this study focuses on cosmological aspects only. We emphasize that this is not the only approach to cosmological perturbations in LQC: the dressed metric (see, {\it e.g.} \cite{Agullo:2012fc}) and hybrid quantization schemes (see, {\it e.g.} \cite{Fernandez-Mendez:2012poe}) also have their benefits (and are related one with the other \cite{Li:2022evi}). \\

The use of generalized holonomy corrections has recently received important attention \cite{Han:2017wmt,Renevey:2021tmh,DeSousa:2022rep,Li:2023axl}, in particular because of remarks made in \cite{Perez:2005fn,Vandersloot:2005kh,BenAchour:2016ajk,Amadei:2022zwp}. It allows one to embed the entire question within a larger framework of generally covariant modifications of general relativity.


    \subsection{Variables and effective corrections of interest}
        \subsubsection{Reduction to the FLRW symmetries and perturbed spacetime}

Let us work with the spacetime manifold $\mathcal{M}$ expressed as  $\mathcal{M}=\mathbb{R}\times\Sigma$. 
The line element reads,
\begin{equation}
ds^2 = -N^2 d\eta^2 + q_{ab}\biggl(N^a d\eta + dx^a\biggr)\biggl(N^b d\eta + dx^b\biggr).
\label{eq:adm_le}
\end{equation}
This introduces three fundamental geometric objects: the lapse function $N$, controlling proper time evolution, the shift vector $N^a$ describing frame dragging effects, and the spatial metric $q_{ab}$ characterizing geometry on constant-time slices $\Sigma$. 

When setting $q_{ab}=a^2(\eta)\delta_{ab}$, $N=a(\eta)$, and $N^a=0$, one is led to the symmetry-reduced FLRW metric,
\begin{equation}
   ds^2 = a^2(\eta)\biggl(-d\eta^2 + \delta_{ab}dx^a dx^b\biggr).
   \label{eq:flrw_le}
\end{equation}
\\

Instead of the spatial metric, LQC relies on the densitized triads $E^a_i$:
\begin{equation}
E^a_i = a^2(\eta)\delta^a_i \equiv \mathfrak{p}(\eta)\delta^a_i.
\end{equation}
In general, triads relate to the spatial metric through $E^a_i E^b_j \delta^{ij} = |\text{det}\, q|q^{ab}$. The variable $\mathfrak{p}$ represents the (squared) scale factor, that is the only dynamical variable of Eq. (\ref{eq:flrw_le}). \\

To perform the canonical analysis of the system, one needs the conjugate variables.
Using the extrinsic curvature tensor,
\begin{equation}
    K_{ab} = \bigl(2N\bigr)^{-1}\biggl(\partial_\eta q_{ab} - 2D_{(a}N_{b)}\biggr),
\end{equation}
they are defined as,
\begin{equation}
    K^i_a = \frac{E^{bi}}{\sqrt{|\det E|}}K_{ab} = \frac{\partial_\eta\mathfrak{p}}{2\mathfrak{p}}\delta^i_a \equiv \mathfrak{c}(\eta)\delta^i_a.
\end{equation}
The variable $\mathfrak{c}$ thus encodes information about the rate of expansion, that is the Hubble parameter $H$. 
The symplectic structure is given by the Poisson bracket:

\begin{equation}
\biggl\{K^i_a(\mathbf{x}), E^b_j(\mathbf{y})\biggr\} = \kappa\delta^i_j\delta^b_a\delta(\mathbf{x}-\mathbf{y}),
\end{equation}

where $\kappa$ is the gravitational  constant. For the homogeneous background variables, this translates to
\begin{equation}\bigl\{\mathfrak{c},\mathfrak{p}\bigr\} = \frac{\kappa}{3\mathcal{V}},
\end{equation}
where $\mathcal{V}$ represents an arbitrary fiducial cell volume introduced to regularize spatial integrals. The Ashtekar connection is defined as
\begin{equation}
    A^i_a = \Gamma^i_a + \gamma K^i_a,
\end{equation}

combining the spin connection $\Gamma^i_a$ with the extrinsic curvature component $K^i_a$, $\gamma$ being the Barbero-Immirzi parameter. 
The associated Poisson bracket reads,
\begin{equation}
\biggl\{A^i_a(\mathbf{x}), E^b_j(\mathbf{y})\biggr\} = \gamma\kappa\delta^i_j\delta^b_a\delta(\mathbf{x}-\mathbf{y}).
\end{equation}
Beyond geometry, one has to include matter. The simplest case is a scalar field $\phi$ with the action:
\begin{equation}
    S_{\frak{m}} = \int d^4x\sqrt{\bigl|\text{det}\,g\bigr|}\biggl[\frac{1}{2}\bigl(\nabla_\mu\phi\bigr)\bigl(\nabla^\mu\phi\bigr) - V[\phi]\biggr],
\end{equation}
where $V[\phi]$ is the potential. The   momentum conjugate to $\phi$ is $\pi$, such that
\begin{equation}
\biggl\{\phi(\mathbf{x}), \pi(\mathbf{y})\biggr\} = \delta(\mathbf{x}-\mathbf{y}).    
\end{equation}

In this work, we shall focus on small inhomogeneities.
The study of cosmological perturbations serves two critical functions in cosmology: it enables the calculation of power spectra - important for phenomenology - and it reveals deep spacetime properties that remain hidden at the homogeneous level. 

The evolution equation for any phase space variable $f$ is $\partial_t f = \{f, H\}$. 
To obtain the linear dynamics of the cosmological perturbations, one therefore needs to perform a second-order expansion of the Hamiltonian.

The lapse and shift functions are decomposed as,
\begin{equation}
    N = \mathbf{N} + \delta N, \quad \text{ and } \quad N^a = \mathbf{N}^a + \delta N^a.
\end{equation}

The gravitational phase space variables are expressed as,
\begin{equation}
    E^a_i = \frak{p} \delta^a_i + \delta E^a_i  \quad \text{ and } \quad K^i_a = \frak{c} \delta^i_a + \delta K^i_a,
\end{equation}

The matter is written as,
\begin{equation}
    \phi = \bm{\phi} + \delta \phi  \quad \text{ and } \quad \pi = \bm{\pi} + \delta \pi.
\end{equation}

\subsubsection{Generalized holonomy corrections}


In the approach used here, curvature is reexpressed as holonomies over a graph.
At the effective level, this is usually implemented by a modification of the symmetry-reduced curvature through the replacement,
\begin{equation}
\mathfrak{c} \rightarrow \frac{\sin(\delta\mathfrak{c})}{\delta},
\end{equation}
where the parameter $\delta(\frak{p})$ is chosen accordingly to a specific  scheme. \\

Substantial ambiguities, however, remain at the level of the quantum dynamics. As explained in details in our previous work (see \cite{DeSousa:2024icf,Han:2017wmt,Renevey:2021tmh}  and reference therein) this is accounted for by using a generic replacement,
\begin{equation}
    \mathfrak{c} \rightarrow g(\mathfrak{c}, \mathfrak{p}).
\end{equation}
This function must fulfill $g(\mathfrak{c}, \mathfrak{p})\rightarrow \mathfrak{c}$ in the low-curvature limit, while respecting the underlying quantum structure.\\ 

This replacement will be used for the background \textit{only}. Another one will be implemented for the perturbative expansion of the constraints, {\it i.e.},
\begin{equation}
    \mathfrak{c} \rightarrow \tilde{g}(\mathfrak{c}, \mathfrak{p}).
\end{equation}
This dual-function approach helps identify the influence of quantum corrections at different structural levels. 
We have previously shown \cite{DeSousa:2024icf} that correcting the perturbations has no influence at all on the observable quantities. 
We shall show in the next sections that when the assumption on the functional dependence of the counterterms is relaxed, this conclusion does not hold anymore and things get trickier. 
    \subsection{Constraints and counterterms}
        \subsubsection{Hamiltonian constraint}
            The full diffeomorphism invariance of general relativity is captured by the Hamiltonian and diffeomorphism constraints, which both include contributions from the geometric and matter sectors. Due to the use of Ashtekar's variables, an extra constraint -- known as the Gauss constraint -- arises to account for the internal gauge freedom associated with the orientation of the triad fields. 
The geometrical contribution to the Hamiltonian constraint is expressed as,
\begin{equation}
  \mathbb{H}_{\mathfrak{g}}[N] = \bigl( 2 \kappa \bigr)^{-1}\! \int \! \! d\mathbf{x} \,  N \frac{E^c_j E^d_k}{\sqrt{\bigl| \text{det} E \bigr|}} \biggl[ \epsilon_i^{jk} F^i_{cd} - 2 \left( 1+ \gamma^2 \right) K^j_{[c}K^k_{d]}\biggr], \label{eq:HGeo}  
\end{equation}
where the field strength $F^i_{ab}$ of the Ashtekar's connection $A^i_a$ is defined by
\begin{equation}
    F^i_{ab} = 2 \partial_{[a} A^i_{b]} + \epsilon^{i}_{jk} A^j_a A^k_b.
\end{equation}
The matter contribution is expressed as
\begin{equation}
    \mathbb{H}_{\mathfrak{m}} =\int \! \! d\mathbf{x} \, N \left[ \frac{\pi_\phi^2}{2\sqrt{\bigl| \text{det} E \bigr|}}+\frac{E^a_i E^b_j }{2\sqrt{\bigl| \text{det} E \bigr|}}\delta^{ij}\partial_a \phi \partial_b \phi + \sqrt{\bigl| \text{det} E \bigr|} V(\phi)\right]. \label{eq:HmDef}
\end{equation}

The Hamiltonian density constraint perturbed at second order is written as the sum of a background part and perturbative expansions for readiness. First, the geometrical contribution to the Hamiltonian constraint is decomposed as a background term,
\begin{equation}
    \mathcal{H}_\frak{g}^{(0)} = - 6 \sqrt{\frak{p}} \frak{c}^2,
\end{equation}
where $X^{(n)}$ stands for the perturbed expression of the quantity $X$ at the $n$th order, a first-order term,
\begin{equation}
    \mathcal{H}_\frak{g}^{(1)} = - 4 \sqrt{\frak{p}}\biggl[1+\alpha_1^{(0)} \biggr] \delta K^b_b - \frac{1}{\sqrt{\frak{p}}}\biggl[\frak{c}^2+\alpha_2^{(0)} \biggr] \delta E^b_b + \frac{2}{\sqrt{\frak{p}}} \biggl[1+\alpha_3^{(0)} \biggr]\partial_a \partial^i \delta E^a_i,
\end{equation}
and a second-order term,
\begin{subequations}
    \begin{align}
    \mathcal{H}_\frak{g}^{(2)} = &\sqrt{\frak{p}} \biggl[1+\alpha_4^{(0)} \biggr]\delta K^a_b \delta K^b_a - \sqrt{\frak{p}} \biggl[1+\alpha_5^{(0)} \biggr](\delta K^b_b) ^2 \\ 
    &- 2 \frac{1}{\sqrt{\frak{p}}} \biggl[\frak{c}+\alpha_6^{(0)} \biggr] \delta K^i_a \delta E^a_i -\frac12 \frac{1}{ \frak{p}^{3/2}}  \biggl[\frak{c}^2+\alpha_7^{(0)} \biggr]  \delta E^a_b \delta E^b_a \\
    &- \frac14 \frac{1}{ \frak{p}^{3/2}} \biggl[\frak{c}^2+\alpha_8^{(0)} \biggr] (\delta E^b_b)^2 + \frac{1}{\frak{p}^{3/2}} \mathcal{Z}_{ab}^{cidj} \biggl[1+\alpha_9^{(0)} \biggr]\bigl( \partial_c \delta E^a_i \bigr) \bigl( \partial_d \delta E^b_j \bigr).
\end{align}
\end{subequations}

In the above we have introduced $\delta K^{b}_{b} \equiv \delta ^b_i \delta K^i_b$ and $\delta E^{b}_{b} \equiv \delta ^i_b \delta E^b_i$, as well as $\mathcal{Z}_{ab}^{cidj}$ defined such that,
\begin{equation}
     \mathcal{Z}_{ab}^{cidj} \equiv \frac14 \epsilon^{ef}_k \epsilon^k_{mn} \mathcal{X}^{mjd}_{be} \mathcal{X}^{nic}_{af} - \epsilon^{ie}_k \mathcal{X}^{kjd}_{bd} \delta^c_a - \epsilon^{ci}_k \mathcal{X}^{kjd}_{ba} + \frac12 \delta^i_a \epsilon^{ce}_k \mathcal{X}^{kjd}_{be},
\end{equation}
with $\epsilon^{ij}_k$ the Levi-Civita symbol and
\begin{equation}
    \mathcal{X}^{ijb}_{ca} \equiv \epsilon^{ij}_c \delta^b_a - \epsilon^{ib}_c \delta^j_a + \epsilon^{ijb} \delta_{ca} + \epsilon^{ib}_a \delta^j_c.
\end{equation}

The matter contribution to the Hamiltonian constraint is written as
\begin{equation}
    \mathcal{H}_{\mathfrak{m}} = \mathcal{H}_{\frak{m}, \pi}+\mathcal{H}_{\frak{m}, \nabla}+\mathcal{H}_{\frak{m}, \phi},
\end{equation}
with
\begin{equation}
    \mathcal{H}_{\frak{m}, \pi} \defeq \frac{\pi_\phi^2}{2\sqrt{\bigl| \text{det} E \bigr|}}, \quad \mathcal{H}_{\frak{m}, \nabla} \defeq \frac{E^a_i E^b_j }{2\sqrt{\bigl| \text{det} E \bigr|}}\delta^{ij}\partial_a \phi \partial_b \phi, \quad \text{ and } \quad \mathcal{H}_{\frak{m}, \phi} \defeq \sqrt{\bigl| \text{det} E \bigr|} V(\phi). \label{eq:ScalarMatterDensityDef}
\end{equation}
The perturbative expansions are therefore given, at the background level, by,
\begin{equation}
    \mathcal{H}_{\frak{m}, \pi}^{(0)} = \frac{\bm{\pi}^2}{2 \frak{p}^{3/2}}, \quad \mathcal{H}_{\frak{m}, \nabla}^{(0)} = 0, \quad \text{ and } \quad \mathcal{H}_{\frak{m}, \phi}^{(0)} = \frak{p}^{3/2} V[\bm{\phi}]. \label{eq:FirstOrderDensityMatterScal}
\end{equation}
At the first order, it becomes,
\begin{align}
    \mathcal{H}_{\frak{m}, \pi}^{(1)} &= \frac{\bm{\pi}}{\frak{p}^{3/2}}  \biggl[1+\beta_1^{(0)} \biggr]\delta \pi - \frac{\bm{\pi}^2}{4 \frak{p}^{5/2}}  \biggl[1+\beta_2^{(0)} \biggr] \delta E^b_b, \\[1.2em]
    \mathcal{H}_{\frak{m}, \phi}^{(1)} &= \frak{p}^{3/2} \bigl(\partial_\phi V\bigr)  \biggl[1+\beta_3^{(0)} \biggr] \delta \phi +  \frac12\frak{p}^{1/2} \, V \biggl[1+\beta_4^{(0)} \biggr] \delta E^b_b, \\[1.2em]
    \mathcal{H}_{\frak{m}, \nabla}^{(1)} &= 0.
\end{align}
The second-order reads,
\begin{subequations}
    \begin{align}
        \mathcal{H}_{\frak{m}, \pi}^{(2)} &= \frac{1}{2} \frak{p}^{-3/2} \biggl[1+\beta_5^{(0)} \biggr] \bigl(\delta \pi\bigr)^2  - \frac12 \frak{p}^{5/2} \bm{\pi}  \biggl[1+\beta_6^{(0)} \biggr]  \delta \pi \delta E^b_b \notag \\ 
        &+ \frac{1}{16}\frak{p}^{-7/2} \bm{\pi}^2  \biggl[1+\beta_9^{(0)} \biggr] (\delta E^b_b)^2 + \frac{1}{8}\frak{p}^{-7/2} \bm{\pi}^2 \biggl[1+\beta_{10}^{(0)} \biggr] \delta E^a_b \delta E^b_a, \\[1.2em]
        \
        \mathcal{H}_{\frak{m}, \phi}^{(2)} &= \frac12 \frak{p}^{3/2} \bigl( \partial_\phi^2 V \bigr) \biggl[1+\beta_7^{(0)} \biggr] \bigl(\delta \phi\bigr)^2 +\frac12 \frak{p}^{1/2} \bigl( \partial_\phi V \bigr)  \biggl[1+\beta_8^{(0)} \biggr] \delta \phi \, \delta E^b_b \notag \\
        &+ \frac18 \frak{p}^{-1/2} \, V \biggl[1+\beta_{11}^{(0)} \biggr] (\delta E^b_b)^2 - \frac14 \frak{p}^{-1/2} \, V  \biggl[1+\beta_{12}^{(0)} \biggr] \delta E^a_b \delta E^b_a \\[1.2em]
        \
        \mathcal{H}_{\frak{m}, \nabla}^{(2)} &=\frac12 \mathfrak{p}^{1/2} \biggl[1+\beta_{13}^{(0)} \biggr] \delta^{ab} \,  \bigl(\partial_a \delta \phi\bigr) \bigl( \partial_b \delta \phi\bigr).
    \end{align}
\end{subequations}
The functions $\alpha_i^{(0)}$ and $\beta_j^{(0)}$ are usually called counterterms as they were initially introduced to cancel the anomalies generated by the holonomy correction. Although we shall keep this wording, it is worth emphasizing that, in light of recent works, they should more appropriately be seen as additional terms allowing -- together with the holonomy correction -- for a consistent (that is first class) generalization of the gravitational theory. In previous works, terms were assumed to depend only on the geometric background components of the phase space, that is on $(\mathfrak{c}, \mathfrak{p})$. While computationally convenient, this assumption can be questioned. 
In this work, we take an initial step into this unexplored territory by extending the functional dependence of the counterterms, using the fields $(K^i_a, E^a_i)$.
        \subsubsection{Diffeomorphism constraint}
The geometrical contribution to the constraint is given by,
\begin{equation}
    \mathbb{D}_{\mathfrak{g}}\defeq \mathbb{D}_{\mathfrak{g}}[N^a] = \bigl( \kappa \gamma \bigr)^{-1}\!\int \! \! d\mathbf{x} \, N^a \mathcal{D}_{a}^{\frak{g}}=\bigl( \kappa \gamma \bigr)^{-1}\!\int \! \! d\mathbf{x} \, N^a\biggl[ \left( \partial_a A^j_b - \partial_b A^j_a \right) E^b_j - A^j_a \partial_b E^b_j \biggr].
\end{equation}
The perturbed density writes
\begin{equation}
    \mathcal{D}_a^{\frak{g}(1)}=\gamma \biggl[\mathfrak{p}\biggl(\partial_a\delta K^b_b-\partial_i \delta K^i_a\biggr)-\mathfrak{c} \delta^j_a \partial_b \delta E^b_j\biggr],
\end{equation}

such that

\begin{equation}
    \mathbb{D}_{\mathfrak{g}} = \bigl( \kappa \gamma \bigr)^{-1} \!\int \! \! d\mathbf{x} \, \delta N^a \mathcal{D}_a^{\frak{g}(1)}.
\end{equation}

Unlike the Hamiltonian constraint, due to the symmetries of the FLRW spacetime, there is no zeroth order contribution. \\


The matter part of the constraint is given by,
\begin{equation}
    \mathbb{D}_{\mathfrak{m}} \defeq \int \! \! d\mathbf{x} \, N^a \pi_{\phi} \bigl(\partial_a \phi \bigr), \label{eq:DmDef}
\end{equation}
and the perturbative expansion reads
\begin{equation}
    \mathcal{D}_{a}^{\mathfrak{m}(1)} = \boldsymbol{\pi} \bigl(\partial_a \delta \phi \bigr).
\end{equation}
Again, there is no zeroth order contribution.
        \subsubsection{Gauss constraint}
The internal rotational symmetry of the triads is captured by the Gauss constraint, expressed as
\begin{equation}
    \mathbb{G} \defeq \mathbb{G}[\Lambda^i] = \bigl( \kappa \gamma \bigr)^{-1}\!\int \! \! d\mathbf{x} \, \Lambda^i \mathcal{G}_i=\bigl( \kappa \gamma \bigr)^{-1}\!\int \! \! d\mathbf{x} \, \Lambda^i \biggl[ \partial_a E^a_i + \epsilon^l_{\, ik} A^k_a E^a_l \biggr],
\end{equation}

which perturbative expansion at second-order is

\begin{equation}
    \mathbb{G} = \bigl( \kappa \gamma \bigr)^{-1}\!\int \! \! d\mathbf{x} \, \delta \Lambda^i \mathcal{G}_i^{(1)},
    \end{equation}

with

\begin{equation}
    \mathcal{G}_i^{(1)}= \gamma \biggl[ \mathfrak{p} \, \epsilon_{ij}^a \delta K^j_a + \mathfrak{c} \, \epsilon_{ia}^j \delta E^a_j \biggr].
\end{equation}
As for the diffeomorphism constraint, there is no zeroth order term in the perturbative expansion. Another similarity between the diffeomorphism and Gauss contraints when perturbed at second order is the absence of terms of the kind $N^{a} \mathcal{D}_{a}^{(2)}$ or $\Lambda^{i} \mathcal{G}_{i}^{(2)}$. This is a consequence of the gauge choices performed at the background level. More details can be found in \cite{DeSousa:2024icf}.
\section{Extension of the functional dependence to $(K^i_a, E^a_i)$}
\subsection{Counterterms expansion}
    As outlined previously, analyses of the effective quantum corrections to the constraint algebra have been, to date, carried out while restricting the counter-term dependence to the background geometrical phase-space variables,  $(\mathfrak{c}, \mathfrak{p})$. While this makes sense given the tedious work required for computing the full constraint algebra, studies within this framework have demonstrated that the induced deformation -- together with the associated change of the space-time signature -- is intimately tied to these counterterms. This might leave imprints in the cosmological power spectra.\\

However, narrowing the counter-term dependence to background variables alone may obscure deeper physical insights. This is why we now broaden the functional dependence to include the full phase-space variables $\bigl(K^i_a, E^a_i\bigr)$. It could also be interesting to use $A^i_a$ instead of $K^i_a$ but we leave this hypothesis for another work. We shall now show that the extension from background variables to $\bigl(K^i_a, E^a_i\bigr)$ yields results that significantly differ from those found in the usual literature.

Thanks to the linearity of the Poisson bracket, future works going beyond this article should be able to build upon the calculations presented here.\\

Let us now write the counterterms as
\begin{equation}
    \alpha_i(K^i_a, E^a_i) = \alpha^{(0)}_i + \alpha^{(1)}_i + \alpha^{(2)}_i
\end{equation}
and
\begin{equation}
    \beta_j(K^i_a, E^a_i) = \beta^{(0)}_j + \beta^{(1)}_j + \beta^{(2)}_j.
\end{equation}
The development can be truncated at the second order as the constraint algebra is studied at this same order. Additionally, a brief inspection of the corrected constraints reveals that only \(\alpha^{(1)}_i\) and \(\beta^{(1)}_j\) are relevant for the present study (\(\alpha^{(0)}_i\) and \(\beta^{(0)}_j\) have already been thoroughly examined in previous works and, as the counterterms always appear multiplied by a density, the expansion at the second order leads to third order terms that are not considered here). The first-order terms are given by:
\begin{equation}
    \alpha^{(1)}_i \defeq \delta E^b_b\bigl(\partial_\mathfrak{p} \alpha_i\bigr)\bigr|_{\mathfrak{c}, \frak{p}}  + \delta K^b_b \bigl(\partial_\mathfrak{c} \alpha_i\bigr)\bigr|_{\mathfrak{c}, \frak{p}}
\end{equation}
and,
\begin{equation}
    \beta^{(1)}_j \defeq \delta E^b_b\bigl(\partial_\mathfrak{p} \beta_j\bigr)\bigr|_{\mathfrak{c}, \frak{p}}  + \delta K^b_b \bigl(\partial_\mathfrak{c} \beta_j\bigr)\bigr|_{\mathfrak{c}, \frak{p}}.
\end{equation}
To make things as clear as possible, we introduce a specific notation to discriminate between new terms coming from the extension of the functional dependence of the counterterms and those that were already present in previous studies (see, {\it e.g.} \cite{DeSousa:2024icf,Renevey:2021tmh}). We write as $\overline{X}$ the extension of the quantity $X$ due to the new  functional dependence of counterterms. This leads to,
\begin{subequations}
\begin{align}
    \overline{\mathcal{H}}_{\mathfrak{g}} &\defeq - 4 \sqrt{\frak{p}} \biggl[\bigl(\pp \alpha_1 \bigr)\tr{E} \tr{K} +\bigl(\pc \alpha_1 \bigr)\bigl( \tr{K} \bigr)^2 \biggr] \\[1.05em]
    &- \frac{1}{\sqrt{\frak{p}}} \biggl[ \bigl(\pc \alpha_2\bigr) \tr{E} \tr{K} + \bigl( \pp \alpha_2 \bigr) \bigl( \tr{E} \bigr)^2 \biggr] \\[1.05em]
    &+ \frac{2}{\sqrt{\frak{p}}} \p_a \p^i \delta E^a_i \biggl[ \bigl(\pp \alpha_3 \bigr) \tr{E} + \bigl(\pc \alpha_3 \bigr) \tr{K}\biggr]
\end{align}
\end{subequations}
and
\begin{subequations}
    \begin{align}
        \overline{\mathcal{H}}_{\mathfrak{m}} &\defeq \frac{\bm{\pi}}{\frak{p}^{3/2}}  \biggl[\bigl(\partial_\mathfrak{p} \beta_1\bigr) \, \delta E^b_b \, \delta \pi + \bigl(\partial_\mathfrak{c} \beta_1\bigr) \, \delta K^b_b \, \delta \pi\biggr] \\[1.05em]
        &- \frac{\bm{\pi}^2}{4 \frak{p}^{5/2}}  \biggl[\bigl(\partial_\mathfrak{p} \beta_2\bigr) \, \bigl(\delta E^b_b\bigr)^2 + \bigl(\partial_\mathfrak{c} \beta_2\bigr) \, \delta K^b_b \, \delta E^b_b \biggr]  \\[1.05em]
        &+ \frak{p}^{3/2} \bigl(\partial_\phi V\bigr)  \biggl[\bigl(\partial_\mathfrak{p} \beta_3\bigr) \, \delta E^b_b \, \delta \phi + \bigl(\partial_\mathfrak{c} \beta_3\bigr) \, \delta K^b_b \, \delta \phi\biggr] \\[1.05em]
        &+  \frac12\frak{p}^{1/2} \, V \biggl[\bigl(\partial_\mathfrak{p} \beta_4\bigr) \, \bigl(\delta E^b_b \bigr)^2 + \bigl(\partial_\mathfrak{c} \beta_4\bigr) \, \delta K^b_b \, \delta E^b_b\biggr], 
    \end{align}
\end{subequations}
where, for readability, we have hidden the background evaluation of the derivatives of the counterterms. \\

This expansion has significant implications for the structure of the constraint algebra, making many of the restrictions established in previous works questionable. Consequently, in the following Sections, we shall explicitly compute the anomalous constraint algebra and search for a consistent solution that yields a first-class system.
\subsection{Brackets computation}
    \subsubsection{Bracket $\left\{ \mathbb{H}, \mathbb{G} \right\}$}
The bracket needs to be extended when compared to the usual case $\alpha_i(\frak{c}, \frak{p})$ and $\beta_j(\mathfrak{c}, \mathfrak{p})$. 
One should now consider,
\begin{equation}
    \biggl\{ \mathbb{H}, \mathbb{G}\biggr\} \longrightarrow \biggl\{ \mathbb{H}, \mathbb{G} \biggr\} + \biggl\{ \overline{\mathbb{H}}, \mathbb{G} \biggr\}.
\end{equation}
As the Gauss constraint $\mathbb{G}$ does not depend on the matter sector,
one can reduce the number of brackets to compute. In particular, for the geometrical sector,
\begin{equation}
    \biggl\{ \overline{\mathbb{H}}_\frak{g}, \mathbb{G}\biggr\}_{\bm{\phi}, \bm{\pi}} = 0 \quad \text{ and } \quad \biggl\{ \overline{\mathbb{H}}_\frak{g}, \mathbb{G}\biggr\}_{\delta \phi, \delta \pi} = 0,
\end{equation}
and, symmetrically, for the matter sector,
\begin{equation}
    \biggl\{ \overline{\mathbb{H}}_\frak{m}, \mathbb{G}\biggr\}_{\bm{\phi}, \bm{\pi}} = 0 \quad \text{ and } \quad \biggl\{ \overline{\mathbb{H}}_\frak{m}, \mathbb{G}\biggr\}_{\delta \phi, \delta \pi} = 0.
\end{equation}
Moreover, derivatives of the Gauss constraints with respect to background quantities lead to second-order quantities in perturbations of the canonical variables. Hence, 
the only relevant term is,
\begin{equation}
    \biggl\{ \overline{\mathbb{H}}, \mathbb{G}\biggr\}_{\frak{c}, \frak{p}} = \biggl\{ \overline{\mathbb{H}}^{(0)}, \mathbb{G}\biggr\}_{\frak{c}, \frak{p}} + o(\delta\delta).
\end{equation}
Nonetheless, as explained in the preceding Sections, extending the functional dependence of the counterterms leads to new terms at the second order.
Therefore,
\begin{equation}
    \biggl\{ \overline{\mathbb{H}}, \mathbb{G}\biggr\}_{\frak{c}, \frak{p}} = o(\delta \delta).
\end{equation}
This simplifies the computation a lot. In particular, only the bracket on the perturbed geometrical phase space leads to nontrivial contributions. In addition,
\begin{equation}
    \biggl\{ \overline{\mathbb{H}}, \mathbb{G}\biggr\}_{\delta E, \delta K} = 0.
    \label{eq:HbarG}
\end{equation}
Since $\overline{\mathbb{H}}$ contains solely traces of the perturbations $\delta E$ and $\delta K$, functional derivatives with respect to those quantities lead to Kronecker deltas only. This, together with the antisymmetry of the Levi-Civita tensor, 
ensures Eq. (\ref{eq:HbarG}). It can therefore be concluded that
\begin{equation}
    \biggl\{ \overline{\mathbb{H}}, \mathbb{G} \biggr\} = 0.
\end{equation}
In other words, the extension of the functional dependence of the counterterms leads neither to new insights about the structure of the algebra nor to restrictions for this particular bracket. This is quite remarkable since, as discussed in \cite{DeSousa:2024icf}, anomalies coming from  \iPB{\mathbb{H}}{\mathbb{G}} are fully related to the anomalies of \iPB{\mathbb{H}}{\mathbb{D}}, thus leading to the conclusion that \iPB{\mathbb{H}}{\mathbb{G}} does not provide anything new. However, the diffeomorphism constraint $\mathbb{D}$ does not share the property of anti-symmetry of $\mathbb{G}$. We therefore expect that this subconclusion does not hold if the functional dependence of the counterterms is extended to $\bigl( K^i_a, E^a_i \bigr)$.
    \subsubsection{Bracket $\left\{ \mathbb{H}, \mathbb{D} \right\}$}
        For the \iPB{\mathbb{H}}{\mathbb{D}} bracket, we proceed in the same manner as for \iPB{\mathbb{H}}{\mathbb{G}}. Extending the functional dependence of the counterterms to $\bigl(K^i_a, E^a_i\bigr)$ leads to,
\begin{equation}
    \biggl\{ \mathbb{H}, \mathbb{D} \biggr\} \stackrel{\alpha(E, K)}{\longrightarrow} \biggl\{ \mathbb{H}, \mathbb{D} \biggr\} + \biggl\{ \overline{\mathbb{H}}, \mathbb{D} \biggr\}.
\end{equation}
The new term can be expressed as
\begin{equation}
    \biggl\{ \overline{\mathbb{H}}, \mathbb{D}\biggr\} =\biggl\{ \overline{\mathbb{H}}_\frak{g}, \mathbb{D}_\frak{g}\biggr\} +\biggl\{ \overline{\mathbb{H}}_\frak{m}, \mathbb{D}_\frak{m}\biggr\} +\biggl\{ \overline{\mathbb{H}}_\frak{g}, \mathbb{D}_\frak{m}\biggr\} +\biggl\{ \overline{\mathbb{H}}_\frak{m}, \mathbb{D}_\frak{g}\biggr\}.
\end{equation}
Some tricks discussed in \cite{DeSousa:2024icf} can be reused here. In particular, as neither $\mathbb{H}_\frak{g}$ (and its counterterms) nor $\mathbb{D}_\frak{g}$ depend on the matter sector, one has
\begin{equation}
    \biggl\{ \overline{\mathbb{H}}_\frak{g}, \mathbb{D}_\frak{g} \biggr\}_{\bm{\phi}, \bm{\pi}}=0 \quad \text{ and } \quad \biggl\{ \overline{\mathbb{H}}_\frak{g}, \mathbb{D}_\frak{g} \biggr\}_{\delta \phi, \delta \pi}=0,
\end{equation}
together with
\begin{equation}
    \biggl\{ \overline{\mathbb{H}}_\frak{m}, \mathbb{D}_\frak{g} \biggr\}_{\bm{\phi}, \bm{\pi}}=0 \quad \text{ and } \quad \biggl\{ \overline{\mathbb{H}}_\frak{m}, \mathbb{D}_\frak{g} \biggr\}_{\delta \phi, \delta \pi}=0.
\end{equation}
As $\mathbb{D}_\frak{m}$ does also not depend upon the geometrical sector, one obtains
\begin{equation}
    \biggl\{ \overline{\mathbb{H}}_\frak{g}, \mathbb{D}_\frak{m} \biggr\}=0,
\end{equation}
together with
\begin{equation}
     \biggl\{ \overline{\mathbb{H}}_\frak{m}, \mathbb{D}_\frak{m} \biggr\}_{\frak{c}, \frak{p}}=0, \quad \text{ and } \quad  \biggl\{ \overline{\mathbb{H}}_\frak{m}, \mathbb{D}_\frak{m} \biggr\}_{\delta E, \delta K}=0.
\end{equation}
As for the \iPB{\overline{H}}{G} bracket, 
one can state that
\begin{equation}
    \biggl\{ \overline{\mathbb{H}}, \mathbb{D} \biggr\}_{\frak{c}, \frak{p}} = \biggl\{ \overline{\mathbb{H}}^{(0)}, \mathbb{D} \biggr\}_{\frak{c}, \frak{p}} + o(\delta \delta).
\end{equation}
As $\overline{\mathbb{H}}^{(0)} = 0$, we shall discard this bracket. Nonetheless, some parts of the bracket have to be explicitly computed.
Starting with \iPB{\mathbb{H}_\mathfrak{g}}{\mathbb{D}_\mathfrak{g}}, one gets
\begin{subequations}
\begin{align}
    \biggl\{ \overline{\mathbb{H}}_\frak{g}[N], \mathbb{D}_\frak{g}[N^a] \biggr\} &= \kappa^{-1} \! \! \int \! \! d\mathbf{x} \, \mathbf{N} \sqrt{\frak{p}} \, \bigl(\partial_a  \delta N^a \bigr) \delta K^b_b \overline{\mathcal{A}}_1^{\{\mathbb{H}, \mathbb{D}\}} \\
    &+ \kappa^{-1} \! \! \int \! \! d\mathbf{x} \, \frac{\mathbf{N}}{2\sqrt{\frak{p}}} \, \bigl( \partial_a \delta N^a\bigr) \delta E^b_b \overline{\mathcal{A}}_2^{\{\mathbb{H}, \mathbb{D}\}}\\
    &+ \kappa^{-1} \! \! \int \! \! d\mathbf{x} \, \frac{\mathbf{N}}{\sqrt{\frak{p}}} \,\bigl(\partial_a \delta N^a\bigr) \bigl(\partial_c \partial_b \delta E^{bc}\bigr)\overline{\mathcal{A}}_3^{\{\mathbb{H}, \mathbb{D}\}},
\end{align}
\end{subequations}
where the first new anomalies are
\begin{align}
    \overline{\mathcal{A}}_1^{\{\mathbb{H}, \mathbb{D}\}} &= -4 \frak{p} (\partial_\frak{p} \alpha_1) - 4 \frak{c} (\partial_c \alpha_1) - (\partial_\frak{c} \alpha_2), \\
     \overline{\mathcal{A}}_2^{\{\mathbb{H}, \mathbb{D}\}} &= -4 \frak{c} \frak{p} (\partial_\frak{p} \alpha_1) - 4 \frak{p} (\partial_\frak{p} \alpha_2) - \frak{c} (\partial_\frak{c} \alpha_2), \\
     \overline{\mathcal{A}}_3^{\{\mathbb{H}, \mathbb{D}\}} &= 2 \frak{p} (\partial_\frak{p} \alpha_3) + \frak{c} (\partial_\frak{c} \alpha_3).
\end{align}
As expected, by construction, those anomalous terms are related only to derivatives of the counterterms. Interestingly, this implies more stringent restrictions when one imposes the anomaly freedom. This point will be made obvious in the next Sections. \\

The \iPB{\mathbb{H}_\mathfrak{m}}{\mathbb{D}} sub-bracket still remains. It is given by
\begin{subequations}
\begin{align}
    \biggl\{ \overline{\mathbb{H}}_\frak{m}[N], \mathbb{D}_\frak{g}[N^a]+\mathbb{D}_\frak{m}[N^a] \biggr\}&= \intnK \frac{\mathbf{N} \bm{\pi}^2}{4 \frak{p}^{5/2}} \bigl( \partial_a \delta N^a \bigr) \delta E^b_b \overline{\mathcal{A}}_4^{\{\mathbb{H}, \mathbb{D}\}} \\
    &+ \intnK \frac{\mathbf{N} \bm{\pi}^2}{\frak{p}^{3/2}} \bigl( \partial_a \delta N^a \bigr) \delta K^b_b \overline{\mathcal{A}}_5^{\{\mathbb{H}, \mathbb{D}\}} \\
    &+ \intnK \frac{\mathbf{N} \bm{\pi}}{\frak{p}^{3/2}} \bigl(\partial_a \delta N^a\bigr) \delta \pi  \overline{\mathcal{A}}_6^{\{\mathbb{H}, \mathbb{D}\}} \\
    &+ \intnK \frac{\mathbf{N} \sqrt{\frak{p}}}{2} V \bigl( \partial_a \delta N^a \bigr) \delta E^b_b \overline{\mathcal{A}}_7^{\{\mathbb{H}, \mathbb{D}\}} \\
   &+ \intnK \mathbf{N} \frak{p}^{3/2} V \bigl( \partial_a \delta N^a \bigr) \delta K^b_b  \overline{\mathcal{A}}_8^{\{\mathbb{H}, \mathbb{D}\}}  \\
   &+ \intnK \frac{\mathbf{N} \sqrt{\frak{p}}}{2} \bigl( \partial_\phi V \bigr) \bigl( \partial_a \delta N^a \bigr) \delta \phi \overline{\mathcal{A}}_9^{\{\mathbb{H}, \mathbb{D}\}},
\end{align}
\end{subequations}
with the anomalous terms,
\begin{align}
    \overline{\mathcal{A}}_4^{\{\mathbb{H}, \mathbb{D}\}} &= 4 \frak{p} \bigl( \partial_\frak{p} \beta_1 \bigr)  - 4 \frak{p} \pp \beta_2 - \frak{c} \pc \beta_2, \\
     \overline{\mathcal{A}}_5^{\{\mathbb{H}, \mathbb{D}\}} &= \partial_\frak{c} \beta_1 - \frac12 \pc \beta_2, \\
      \overline{\mathcal{A}}_6^{\{\mathbb{H}, \mathbb{D}\}} &= 2 \frak{p} \pp \beta_1 + \frak{c} \pc \beta_1, \\
      \overline{\mathcal{A}}_7^{\{\mathbb{H}, \mathbb{D}\}} &= 4 \frak{p} \pp \beta_4 + \frak{c} \pc \beta_4, \\
      \overline{\mathcal{A}}_8^{\{\mathbb{H}, \mathbb{D}\}} &=\pc \beta_4, \\
      \overline{\mathcal{A}}_9^{\{\mathbb{H}, \mathbb{D}\}}&=4 \frak{p}^2 \pp \beta_3 + 2 \frak{c}\frak{p} \pc \beta_3.
\end{align}

    \subsubsection{Bracket $\left\{ \mathbb{H}, \mathbb{H} \right\}$}
        Finally, let us focus on the most complicated Poisson bracket: $\{\mathbb{H},\mathbb{H}\}$. As the full phase space is involved, this calculation is challenging. 
We proceed as follows:
\begin{align}
     \biggl\{ \mathbb{H}[N_1], \mathbb{H}[N_2]\biggr\} \stackrel{\alpha(E, K)}{\longrightarrow} &\biggl\{ \mathbb{H}[N_1], \mathbb{H}[N_2]\biggr\} + \biggl\{ \overline{\mathbb{H}}[N_1], \overline{\mathbb{H}}[N_2]\biggr\} \notag \\
     &+ \biggl[ \biggl\{ \overline{\mathbb{H}}[N_1], \mathbb{H}[N_2]\biggr\} - \bigl( N_1 \leftrightarrow N_2 \bigr)\biggr].
\end{align}
For readability, we define
\begin{equation}
    \Delta \biggl\{ \overline{\mathbb{H}}, \mathbb{H} \biggr\} \defeq \biggl[ \biggl\{ \overline{\mathbb{H}}[N_1], \mathbb{H}[N_2]\biggr\} - \bigl( N_1 \leftrightarrow N_2 \bigr)\biggr].
\end{equation}
Since the background gauge is fixed due to symmetries, the background pars of $N_1$ and $N_2$ are equal. Therefore, one easily sees that
\begin{equation}
    \biggl\{ \overline{\mathbb{H}}[N_1], \overline{\mathbb{H}}[N_2]\biggr\} = 0.
\end{equation}
This leads to the conclusion that the new terms are, by themselves, irrelevant for this bracket. Their significance is primarily intertwined with the terms examined in previous works. This complicates the resolution but raises interesting new relations. \\

For the geometrical component of the Hamiltonian constraint, one has, 
\begin{subequations}
\begin{align}
    \Delta \biggl\{ \overline{\mathbb{H}}_\frak{g}, \mathbb{H}_\frak{g} \biggr\} &= \intwK \mathbf{N} \Delta \bigl[ \delta N \bigr] \delta K^b_b \, \overline{\mathcal{A}}_1^{\{\mathbb{H}, \mathbb{H}\}} \\
    &+ \intwK \frac{\mathbf{N}}{2 \frak{p}} \Delta \bigl[ \delta N \bigr] \delta E^b_b \, \overline{\mathcal{A}}_2^{\{\mathbb{H}, \mathbb{H}\}} \\
    &+ \intwK \frac{\mathbf{N}}{2 \frak{p}} \bigl(\partial_a \partial^a \Delta \bigl[ \delta N \bigr] \bigr) \delta E^b_b \, \overline{\mathcal{A}}_3^{\{\mathbb{H}, \mathbb{H}\}} \\
    &+ \intwK \mathbf{N} \, \partial_a \partial^a \bigl(\Delta \bigl[ \delta N \bigr] \bigr) \delta K^b_b \, \overline{\mathcal{A}}_4^{\{\mathbb{H}, \mathbb{H}\}} \\
    &+ \intwK \frac{\mathbf{N}}{\frak{p}} \, \Delta \bigl[ \delta N \bigr] \bigl( \partial_b\partial_a \delta E^{ab} \bigr) \, \overline{\mathcal{A}}_5^{\{\mathbb{H}, \mathbb{H}\}} \\
    &+ \intwK \frac{\mathbf{N}}{\frak{p}} \bigl( \partial_a \partial^a \Delta \bigl[ \delta N \bigr] \bigr) \bigl(\partial_c \partial_b \delta E^{bc}\bigr) \, \overline{\mathcal{A}}_6^{\{\mathbb{H}, \mathbb{H}\}}, 
\end{align}
\end{subequations}
where the anomalous terms are
\begin{align}
    \overline{\mathcal{A}}_1^{\{\mathbb{H}, \mathbb{H}\}} &= -12 \frak{p} \bigl( \pp{\alpha_1} \bigr) \biggl[ \tilde{g} + \alpha_1 \biggr] + 6 \bigl( \pc{\alpha_1} \bigr) \biggl[ \tilde{g}^2 + \alpha_2 \biggr] - 3 \bigl( \pc{\alpha_2} \bigr) \biggl[ \tilde{g} + \alpha_1 \biggr], \\
    \overline{\mathcal{A}}_2^{\{\mathbb{H}, \mathbb{H}\}} &= 6 \frak{p} \bigl(\pp{\alpha_1}\bigr) \biggl[ \tilde{g}^2 + \alpha_2 \biggr] - 12 \frak{p} \bigl(\pp{\alpha_2}\bigr) \biggl[ \tilde{g} + \alpha_1 \biggr] + \frac32 \bigl(\pc{\alpha_2}\bigr) \biggl[ \tilde{g}^2 + \alpha_2 \biggr], \\
    \overline{\mathcal{A}}_3^{\{\mathbb{H}, \mathbb{H}\}} &= 4 \frak{p} \bigl( \pp{\alpha_3} \bigr) \biggl[ \tilde{g}+\alpha_1 \biggr] - 4 \frak{p} \bigl( \pp{\alpha_1} \bigr) \biggl[ 1+\alpha_3\biggr] - \bigl( \pc{\alpha_2} \bigr) \biggl[ 1 + \alpha_3 \biggr], \\
    \overline{\mathcal{A}}_4^{\{\mathbb{H}, \mathbb{H}\}} &= 2 \bigl( \pc{\alpha_3} \bigr) \biggl[ \tilde{g} + \alpha_1 \biggr] - 4 \bigl(\pc{\alpha_1}\bigr) \biggl[ 1 + \alpha_3 \biggr], \\
    \overline{\mathcal{A}}_5^{\{\mathbb{H}, \mathbb{H}\}} &= 6 \frak{p} \bigl(\pp{\alpha_3}\bigr) \biggl[ \tilde{g} + \alpha_1 \biggr] - \frac32 \bigl(\pc{\alpha_3}\bigr) \biggl[ \tilde{g}^2 + \alpha_2 \biggr], \\  
    \overline{\mathcal{A}}_6^{\{\mathbb{H}, \mathbb{H}\}} &= \pc{\alpha_3}.
\end{align}
For the cross terms between the geometrical and matter sectors, one gets,
\begin{subequations}
\begin{align}
    \Delta \biggl\{ \overline{\mathbb{H}}_\frak{g}, \mathbb{H}_\frak{m} \biggr\} + \Delta \biggl\{ \overline{\mathbb{H}}_\frak{m}, \mathbb{H}_\frak{g} \biggr\}  &= \intwK \frac{\mathbf{N}}{4}V \Delta \bigl[ \delta N \bigr] \delta E^b_b\, \overline{\mathcal{A}}_7^{\{\mathbb{H}, \mathbb{H}\}} \\
    &+ \intwK \frac{\mathbf{N} \bm{\pi}^2}{8 \frak{p}^3} \Delta \bigl[ \delta N \bigr] \delta E^b_b\,\overline{\mathcal{A}}_8^{\{\mathbb{H}, \mathbb{H}\}} \\
    &+ \intwK \frac{\mathbf{N} \bm{\pi}}{2 \frak{p}^2}\Delta \bigl[ \delta N \bigr] \delta \pi\,\overline{\mathcal{A}}_9^{\{\mathbb{H}, \mathbb{H}\}} \\
    &+ \intwK \mathbf{N} \bigl( \partial_\phi V \bigr) \Delta \bigl[ \delta N \bigr] \delta \phi  \, \overline{\mathcal{A}}_{10}^{\{\mathbb{H}, \mathbb{H}\}} \\
    &+ \intwK \frac{\mathbf{N} \frak{p}}{2} V \Delta \bigl[ \delta N \bigr] \delta K^b_b\,\overline{\mathcal{A}}_{11}^{\{\mathbb{H}, \mathbb{H}\}} \\
    &+ \intwK \frac{\mathbf{N} \bm{\pi}^2}{4 \frak{p}^2} \Delta \bigl[ \delta N \bigr] \delta K^b_b\,\overline{\mathcal{A}}_{12}^{\{\mathbb{H}, \mathbb{H}\}} \\
    &+ \intwK \frac{\mathbf{N}}{2}V\Delta \bigl[ \delta N \bigr] \bigl( \partial_b \partial_a \delta E^{ab} \bigr)\,\overline{\mathcal{A}}_{13}^{\{\mathbb{H}, \mathbb{H}\}} \\
    &+ \intwK \frac{\mathbf{N}\bm{\pi}^2}{4\frak{p}^3}\Delta \bigl[ \delta N \bigr] \bigl( \partial_b \partial_a \delta E^{ab} \bigr)\,\overline{\mathcal{A}}_{14}^{\{\mathbb{H}, \mathbb{H}\}} \\
    &+ \intwK \frac{\mathbf{N} \bm{\pi}}{\frak{p}^2} \bigl( \partial_a \partial^a \Delta \bigl[ \delta N \bigr] \bigr) \delta \pi \,\overline{\mathcal{A}}_{15}^{\{\mathbb{H}, \mathbb{H}\}} \\
    &+ \intwK \frac{\mathbf{N} \bm{\pi}^2}{4 \frak{p}^3}\bigl( \partial_a \partial^a \Delta \bigl[ \delta N \bigr] \bigr) \delta E^b_b \, \overline{\mathcal{A}}_{16}^{\{\mathbb{H}, \mathbb{H}\}} \\
    &+ \intwK \mathbf{N} \frak{p} \bigl( \partial_\phi V \bigr) \bigl( \partial_a \partial^a \Delta \bigl[ \delta N \bigr] \bigr) \delta \phi \, \overline{\mathcal{A}}_{17}^{\{\mathbb{H}, \mathbb{H}\}} \\
    &+ \intwK \frac{\mathbf{N}}{2} V \bigl( \partial_a \partial^a \Delta \bigl[ \delta N \bigr] \bigr) \delta E^b_b \, \overline{\mathcal{A}}_{18}^{\{\mathbb{H}, \mathbb{H}\}},
\end{align}
\end{subequations}
with
\begin{align}
    \overline{\mathcal{A}}_7^{\{\mathbb{H}, \mathbb{H}\}} &= 24 \frak{p} \bigl( \pp{\beta_4} \bigr) \biggl[ \tilde{g} + \alpha_1 \biggr] - 12 \frak{p} \bigl( \pp{\alpha_1} \bigr) \biggl[ 1 + \beta_4 \biggr] - 3 \bigl( \pc{\alpha_2} \bigr) \biggl[ 1 + \beta_4 \biggr] \notag \\ &\qquad - 3 \bigl( \pc{\beta_4} \bigr) \biggl[ \tilde{g}^2 + \alpha_2 \biggr], \\
     \overline{\mathcal{A}}_8^{\{\mathbb{H}, \mathbb{H}\}} &= 12 \frak{p} \bigl( \pp{\alpha_1} \bigr) \biggl[ 1 + \beta_2 \biggr] - 24 \frak{p} \bigl( \pp{\beta_2} \bigr) \biggl[ \tilde{g} + \alpha_1 \biggr] + 3 \bigl( \pc{\alpha_2} \bigr) \biggl[ 1 + \beta_2 \biggr] + 3 \bigl( \pc{\beta_2} \bigr) \biggl[ \tilde{g}^2 + \alpha_2 \biggr], \\
     \overline{\mathcal{A}}_9^{\{\mathbb{H}, \mathbb{H}\}} &= 12 \frak{p} \bigl( \pp{\beta_1} \bigr) \biggl[ \tilde{g} + \alpha_1 \biggr] - 3 \bigl( \pc{\beta_1} \bigr) \biggl[ \tilde{g}^2 + \alpha_2 \biggr], \\
     \overline{\mathcal{A}}_{10}^{\{\mathbb{H}, \mathbb{H}\}} &= 6 \frak{p}^2 \bigl( \pp{\beta_3} \bigr) \biggl[ \tilde{g} + \alpha_1 \biggr] - \frac32 \frak{p} \bigl( \pc{\beta_3} \bigr) \biggl[ \tilde{g}^2 + \alpha_2 \biggr], \\
     \overline{\mathcal{A}}_{11}^{\{\mathbb{H}, \mathbb{H}\}} &= 6 \bigl( \pc{\beta_4} \bigr) \biggl[ \tilde{g} + \alpha_1 \biggr]-12 \bigl( \pc{\alpha_1} \bigr) \biggl[ 1 + \beta_4 \biggr], \\
     \overline{\mathcal{A}}_{12}^{\{\mathbb{H}, \mathbb{H}\}} &= 12 \bigl( \pc{\alpha_1} \bigr) \biggl[ 1 + \beta_2 \biggr] - 6 \bigl( \pc{\beta_2} \bigr) \biggl[ \tilde{g} + \alpha_1 \biggr], \\
     \overline{\mathcal{A}}_{13}^{\{\mathbb{H}, \mathbb{H}\}} &= 3 \bigl( \pc{\alpha_3} \bigr) \biggl[ 1 + \beta_4 \biggr], \\
     \overline{\mathcal{A}}_{14}^{\{\mathbb{H}, \mathbb{H}\}} &= -3 \bigl( \pc{\alpha_3} \bigr) \biggl[ 1 + \beta_2 \biggr], \\
     \overline{\mathcal{A}}_{15}^{\{\mathbb{H}, \mathbb{H}\}} &= \bigl( \pc{\beta_1} \bigr) \biggl[ 1 + \alpha_3 \biggr], \\
     \overline{\mathcal{A}}_{16}^{\{\mathbb{H}, \mathbb{H}\}} &= \bigl( \pc{\beta_2} \bigr) \biggl[ 1 + \alpha_3 \biggr], \\
     \overline{\mathcal{A}}_{17}^{\{\mathbb{H}, \mathbb{H}\}} &= \bigl( \pc{\beta_3} \bigr) \biggl[ 1 + \alpha_3 \biggr], \\
     \overline{\mathcal{A}}_{18}^{\{\mathbb{H}, \mathbb{H}\}} &= \bigl( \pc{\beta_4} \bigr) \biggl[ 1 + \alpha_3 \biggr].
\end{align}
As we shall see later, the requirement that general relativity is recovered in the classical limit imposes here very stringent restrictions.\\

For the matter sector, one gets,
\begin{subequations}
\begin{align}
     \Delta \biggl\{ \overline{\mathbb{H}}_\frak{m}, \mathbb{H}_\frak{m} \biggr\} &= \intnK \frac{\mathbf{N} \bm{\pi}}{2 \frak{p}} \bigl( \partial_\phi V \bigr)  \Delta \bigl[ \delta N \bigr] \delta E^b_b \, \overline{\mathcal{A}}_{19}^{\{\mathbb{H}, \mathbb{H}\}} \\
     &+ \intnK \frac{\mathbf{\mathbf{N} \bm{\pi}}}{2 \frak{p}} V  \Delta \bigl[ \delta N \bigr] \delta \pi\, \overline{\mathcal{A}}_{20}^{\{\mathbb{H}, \mathbb{H}\}} \\
     &+ \intnK \frac{\mathbf{N} \bm{\pi}^3}{4 \frak{p}^4}  \Delta \bigl[ \delta N \bigr]\delta \pi\, \overline{\mathcal{A}}_{21}^{\{\mathbb{H}, \mathbb{H}\}} \\ 
     &+ \intnK \mathbf{N} \bm{\pi} \bigl( \partial_\phi V \bigr)  \Delta \bigl[ \delta N \bigr]\delta K^b_b \, \overline{\mathcal{A}}_{22}^{\{\mathbb{H}, \mathbb{H}\}} \\
     &+ \intnK \frac{\mathbf{N} \bm{\pi}^2}{8 \frak{p}^2} V \Delta \bigl[ \delta N \bigr]\delta E^b_b \, \overline{\mathcal{A}}_{23}^{\{\mathbb{H}, \mathbb{H}\}} \\
     &+ \intnK \frac{\mathbf{N} \bm{\pi}^4}{16 \frak{p}^5}\Delta \bigl[ \delta N \bigr]\delta E^b_b \, \overline{\mathcal{A}}_{24}^{\{\mathbb{H}, \mathbb{H}\}} \\
     &+ \intnK \frac{\mathbf{N} \frak{p}^2}{2} \bigl( \partial_\phi V \bigr) V \Delta \bigl[ \delta N \bigr] \delta \phi  \, \overline{\mathcal{A}}_{25}^{\{\mathbb{H}, \mathbb{H}\}} \\
     &+ \intnK \frac{\mathbf{N} \bm{\pi}^2}{4 \frak{p}} \bigl( \partial_\phi V \bigr) \Delta \bigl[ \delta N \bigr] \delta \phi  \, \overline{\mathcal{A}}_{26}^{\{\mathbb{H}, \mathbb{H}\}} \\
     &+ \intnK \frac{\mathbf{N} \frak{p}}{4} V^2 \Delta \bigl[ \delta N \bigr] \delta E^b_b  \, \overline{\mathcal{A}}_{27}^{\{\mathbb{H}, \mathbb{H}\}}, 
\end{align}
\end{subequations}
with 
\begin{align}
    \overline{\mathcal{A}}_{19}^{\{\mathbb{H}, \mathbb{H}\}} &= 2\frak{p} \bigl( \pp{\beta_3} \bigr) \biggl[ 1 + \beta_1 \biggr] - 2 \frak{p} \bigl( \pp{\beta_1} \bigr) \biggl[ 1 + \beta_3 \biggr], \\
    \overline{\mathcal{A}}_{20}^{\{\mathbb{H}, \mathbb{H}\}} &= 3 \bigl( \pc{\beta_1} \bigr) \biggl[ 1 + \beta_4 \biggr], \\
    \overline{\mathcal{A}}_{21}^{\{\mathbb{H}, \mathbb{H}\}} &= -3  \bigl( \pc{\beta_1} \bigr) \biggl[ 1 + \beta_2 \biggr], \\
    \overline{\mathcal{A}}_{22}^{\{\mathbb{H}, \mathbb{H}\}} &= \bigl( \pc{\beta_3} \bigr) \biggl[ 1 + \beta_1 \biggr] - \bigl( \pc{\beta_1} \bigr) \biggl[ 1 + \beta_3 \biggr], \\
    \overline{\mathcal{A}}_{23}^{\{\mathbb{H}, \mathbb{H}\}} &= -3 \bigl( \pc{\beta_2} \bigr) \biggl[ 1 + \beta_4 \biggr]-3 \bigl( \pc{\beta_4} \bigr) \biggl[ 1 + \beta_2 \biggr], \\
    \overline{\mathcal{A}}_{24}^{\{\mathbb{H}, \mathbb{H}\}} &= 3 \bigl( \pc{\beta_2} \bigr) \biggl[ 1 + \beta_2 \biggr], \\
    \overline{\mathcal{A}}_{25}^{\{\mathbb{H}, \mathbb{H}\}} &= 3 \bigl( \pc{\beta_3} \bigr) \biggl[ 1 + \beta_4 \biggr], \\
    \overline{\mathcal{A}}_{26}^{\{\mathbb{H}, \mathbb{H}\}} &= -3 \bigl( \pc{\beta_3} \bigr) \biggl[ 1 + \beta_2 \biggr], \\
    \overline{\mathcal{A}}_{27}^{\{\mathbb{H}, \mathbb{H}\}} &= 3 \bigl( \pc{\beta_4} \bigr) \biggl[ 1 + \beta_4 \biggr].
\end{align}
Again, the classical limit will lead to important restrictions. 
\subsection{Solution to close the algebra of constraints}
    \subsubsection{Partial solution to the reduced system of anomalies}
Remarkably, some of the terms derived in the previous section are {\it independent} of the anomalies derived in \cite{DeSousa:2024icf}. We shall first focus on those terms. A close examination shows that that  $\overline{\mathcal{A}}_3^{\{\mathbb{H}, \mathbb{D}\}}$, $\overline{\mathcal{A}}_5^{\{\mathbb{H}, \mathbb{D}\}}$, $\overline{\mathcal{A}}_8^{\{\mathbb{H}, \mathbb{D}\}}$, $\overline{\mathcal{A}}_4^{\{\mathbb{H}, \mathbb{H}\}}$, $\overline{\mathcal{A}}_6^{\{\mathbb{H}, \mathbb{H}\}}$, $\overline{\mathcal{A}}_{15}^{\{\mathbb{H}, \mathbb{H}\}}$, $\overline{\mathcal{A}}_{16}^{\{\mathbb{H}, \mathbb{H}\}}$, $\overline{\mathcal{A}}_{17}^{\{\mathbb{H}, \mathbb{H}\}}$, $\overline{\mathcal{A}}_{18}^{\{\mathbb{H}, \mathbb{H}\}}$, $\overline{\mathcal{A}}_{21}^{\{\mathbb{H}, \mathbb{H}\}}$, and $\overline{\mathcal{A}}_{22}^{\{\mathbb{H}, \mathbb{H}\}}$ are purely related to the functional expansion of the counterterms. \\

As $\alpha_3$ cannot be a constant (otherwise the classical limit would not be recovered), one has
\begin{equation}
    \pc{\beta_1}=0, \quad \pc{\beta_2}=0, \quad \pc{\beta_3}=0, \quad \text{and} \quad \pc{\beta_4}=0,
    \label{eq:pcb1234}
\end{equation}
from, respectively, $\overline{\mathcal{A}}_{15}^{\{\mathbb{H}, \mathbb{H}\}}$, $\overline{\mathcal{A}}_{16}^{\{\mathbb{H}, \mathbb{H}\}}$, $\overline{\mathcal{A}}_{17}^{\{\mathbb{H}, \mathbb{H}\}}$, and $\overline{\mathcal{A}}_{18}^{\{\mathbb{H}, \mathbb{H}\}}$. Equation (\ref{eq:pcb1234}), in turns, implies that the conditions
\begin{equation}
    \overline{\mathcal{A}}_{20}^{\{\mathbb{H}, \mathbb{H}\}}=0\,...\,\overline{\mathcal{A}}_{27}^{\{\mathbb{H}, \mathbb{H}\}}=0, \quad \overline{\mathcal{A}}_{5}^{\{\mathbb{H}, \mathbb{D}\}}=0, \quad \text{and} \quad \overline{\mathcal{A}}_{8}^{\{\mathbb{H}, \mathbb{D}\}}=0,
\end{equation}
are automatically satisfied. Thanks to $\overline{\mathcal{A}}_{6}^{\{\mathbb{H}, \mathbb{H}\}}$, one can easily obtain 
\begin{equation}
    \pc \alpha_3 = 0,
\end{equation}
which, in turn, using $\overline{\mathcal{A}}_{3}^{\{\mathbb{H}, \mathbb{D}\}}$, implies
\begin{equation}
    \pp \alpha_3 = 0.
\end{equation}
This shows that 
\begin{equation}
    \alpha_3 = 0.
    \label{eq:alpha3sol}
\end{equation}
This has an important consequence. When the counterterms are functions of the geometrical background fields only, there exists an ambiguity between $\alpha_3$ and $\alpha_9$. The closure of the algebra does not fully determine its structure \cite{DeSousa:2024icf}. This ambiguity is fixed by the procedure followed here, that is once the functional dependence of the counterterms is expanded to the full fields $(K^i_a, E^a_i)$. Implementing Eq.~ (\ref{eq:alpha3sol}) into $\overline{\mathcal{A}}_{4}^{\{\mathbb{H}, \mathbb{H}\}}$, one obtains
\begin{equation}
    \pc{\alpha_1} = 0.
    \label{eq:pca1}
\end{equation}
This implies in turn that $\overline{\mathcal{A}}_{11}^{\{\mathbb{H}, \mathbb{H}\}}$ and $\overline{\mathcal{A}}_{12}^{\{\mathbb{H}, \mathbb{H}\}}$ vanish. Again, using Eq.~(\ref{eq:alpha3sol}), we automatically get rid of  $\overline{\mathcal{A}}_{13}^{\{\mathbb{H}, \mathbb{H}\}}$ and  $\overline{\mathcal{A}}_{14}^{\{\mathbb{H}, \mathbb{H}\}}$. \\

This is all that can be extracted from independent anomalies. The full system now has to be considered.


    \subsubsection{Solution to the complete system of anomalies}
Taking into account results from \cite{DeSousa:2024icf}, one can calculate the Poisson brackets between the \textit{full} constraints. Let us start with the geometrical contribution to \iPB{\mathbb{H}}{\mathbb{D}},
\begin{subequations}
\begin{align}
    \biggl\{\mathbb{H}_\frak{g}[N], \mathbb{D}_\frak{g}[N^a] \biggr\} &= \mathbb{H}_\frak{g}[\delta N \partial_a \delta N^a] \label{eq:HDFull1}\\ 
     &+ \intwK \sqrt{\frak{p}} \, \delta N \bigl( \partial_a  \delta N^a \bigr) \mathcal{A}_1^{{\left\{ \mathbb{H}, \mathbb{D} \right\}}}\\
     &+ \intwK \frac{\mathbf{N}}{2\sqrt{\frak{p}}} \, \bigl(\partial_a \delta N^a\bigr) \delta E^b_b \mathcal{A}_2^{{\left\{ \mathbb{H}, \mathbb{D} \right\}}} \\
     &+ \intwK \mathbf{N} \sqrt{\frak{p}} \,\bigl(\partial_a\delta N^a\bigr) \delta K^b_b \mathcal{A}_3^{{\left\{ \mathbb{H}, \mathbb{D} \right\}}} \\
     &+ \intwK \frac{\mathbf{N}}{2\sqrt{\frak{p}}} \,\bigl(\partial_a \delta N^b\bigr) \delta^i_b \delta E^a_i \mathcal{A}_4^{{\left\{ \mathbb{H}, \mathbb{D} \right\}}} \\
     &+ \intwK \mathbf{N} \sqrt{\frak{p}} \,\bigl(\partial_a \delta N^b\bigr) \delta^a_i \delta K^i_b  \mathcal{A}_5^{{\left\{ \mathbb{H}, \mathbb{D} \right\}}} \label{eq:HDFull2},
\end{align}
\label{eq:HDFull}
\end{subequations}
where the anomalies are, 
\begin{align}
    \mathcal{A}_1^{{\left\{ \mathbb{H}, \mathbb{D} \right\}}} &= 3 g^2 -  \bigl[\tilde{g}^2 + \alpha_2 \bigr] - 2 \frak{c} \bigl[ \tilde{g} + \alpha_1  \bigr], \label{eq:AHD1} \\
    \mathcal{A}_2^{{\left\{ \mathbb{H}, \mathbb{D} \right\}}}&= \alpha_8 - \alpha_7 + \overline{\mathcal{A}}_2^{{\left\{ \mathbb{H}, \mathbb{D} \right\}}},\label{eq:AHD2}\\
    \mathcal{A}_3^{{\left\{ \mathbb{H}, \mathbb{D} \right\}}} &= \frak{c}\bigl[1 + \alpha_5 \bigr] + \bigl[\tilde{g} + \alpha_6 \bigr] - 2 g \partial_\frak{c}g + \overline{\mathcal{A}}_1^{{\left\{ \mathbb{H}, \mathbb{D} \right\}}}, \label{eq:AHD3} \\
    \mathcal{A}_4^{{\left\{ \mathbb{H}, \mathbb{D} \right\}}} &= g^2 - 2 \frak{c}\bigl[ \tilde{g} + \alpha_6 \bigr] + \bigl[\tilde{g}^2 + \alpha_7 \bigr] + 4 \frak{p} g \partial_\frak{p}g, \label{eq:AHD4}  \\
    \mathcal{A}_5^{{\left\{ \mathbb{H}, \mathbb{D} \right\}}} &= \frak{c}\bigl[1 + \alpha_4] + \bigl[\tilde{g}  + \alpha_6\bigr] - 2 g \partial_\frak{c}g.
    \label{eq:AHD5}
\end{align}

This has significant implications for the algebraic structure, as will soon become apparent. For the matter contribution to the Hamiltonian constraint $\mathbb{H}$ and to the diffeomorphism constraint $\mathbb{D}$, one has,
\begin{subequations}
\begin{align}
    \biggl\{\mathbb{H}_\frak{m}[N], \mathbb{D}_\frak{g}[N^a]+\mathbb{D}_\frak{m}[N^a]\biggr\} &= \mathbb{H}_\frak{m}[\delta N \partial_a \delta N^a] \label{eq:HDtot1}\\
    &+ \intnK \frac{\mathbf{N} \bm{\pi}}{\frak{p}^{3/2}} \,  \bigl(\partial_a \delta N^a \bigr) \delta \pi  \mathcal{A}_6^{{\left\{ \mathbb{H}, \mathbb{D} \right\}}} \\
     &+ \intnK \frac{\bm{\pi}^2}{2 \frak{p}^{3/2}} \, \delta N \bigl( \partial_a \delta N^a \bigr)\mathcal{A}_7^{{\left\{ \mathbb{H}, \mathbb{D} \right\}}} \\
     &+ \intnK \frak{p}^{3/2}\, V \, \delta N  \bigl(\partial_a \delta N^a \bigr) \mathcal{A}_8^{{\left\{ \mathbb{H}, \mathbb{D} \right\}}} \\
     &+ \intnK \frac{\mathbf{N} \bm{\pi}^2}{4 \frak{p}^{5/2}} \,  \bigl(\partial_a \delta N^b \bigr) \delta^i_b \delta E^a_i \mathcal{A}_9^{{\left\{ \mathbb{H}, \mathbb{D} \right\}}} \\
     &+ \intnK \frac{\mathbf{N} \sqrt{\frak{p}}}{2} V \,\bigl( \partial_a \delta N^b \bigr) \delta^i_b \delta E^a_i \mathcal{A}_{10}^{{\left\{ \mathbb{H}, \mathbb{D} \right\}}} \\
     &+ \intnK \frac{\mathbf{N} \bm{\pi}}{4 \frak{p}^{5/2}} \,  \bigl(\partial_a \delta N^a \bigr) \delta E^b_b \mathcal{A}_{11}^{{\left\{ \mathbb{H}, \mathbb{D} \right\}}}\\
     &+ \intnK \frac{\mathbf{N} \sqrt{\frak{p}}}{2} \,  \bigl(\partial_\phi V \bigr) \, \bigl(  \partial_a  \delta N^a \bigr)\delta \phi \mathcal{A}_{12}^{{\left\{ \mathbb{H}, \mathbb{D} \right\}}}\\
     &+  \intnK \frac{\mathbf{N} \sqrt{\frak{p}}}{2} \, V  \bigl( \partial_a \delta N^a \bigr) \delta E^b_b \mathcal{A}_{13}^{{\left\{ \mathbb{H}, \mathbb{D} \right\}}}, \label{eq:HDtot2}
\end{align}
\label{eq:HDtot}
\end{subequations}
the anomalies being
\begin{align}
     \mathcal{A}_6^{{\left\{ \mathbb{H}, \mathbb{D} \right\}}} &= \beta_5 - \beta_6 + \overline{\mathcal{A}}_{6}^{{\left\{ \mathbb{H}, \mathbb{D} \right\}}}, \label{eq:AHD6}\\
     \mathcal{A}_7^{{\left\{ \mathbb{H}, \mathbb{D} \right\}}} &= 2 \beta_1 - \beta_2, \label{eq:AHD7}\\
      \mathcal{A}_8^{{\left\{ \mathbb{H}, \mathbb{D} \right\}}} &= \beta_4, \label{eq:AHD8}\\
      \mathcal{A}_9^{{\left\{ \mathbb{H}, \mathbb{D} \right\}}} &= - \beta_{10}, \label{eq:AHD9}\\
      \mathcal{A}_{10}^{{\left\{ \mathbb{H}, \mathbb{D} \right\}}} &= \beta_{12}, \label{eq:AHD10}\\
        \mathcal{A}_{11}^{{\left\{ \mathbb{H}, \mathbb{D} \right\}}} &= \beta_{10} - 2 \beta_6 + \beta_9 + \overline{\mathcal{A}}_{4}^{{\left\{ \mathbb{H}, \mathbb{D} \right\}}}, \label{eq:AHD11}\\
        \mathcal{A}_{12}^{{\left\{ \mathbb{H}, \mathbb{D} \right\}}} &= \beta_8 + \overline{\mathcal{A}}_{9}^{{\left\{ \mathbb{H}, \mathbb{D} \right\}}}, \label{eq:AHD12}\\
        \mathcal{A}_{13}^{{\left\{ \mathbb{H}, \mathbb{D} \right\}}} &= \beta_{11} - \beta_{12} + \overline{\mathcal{A}}_{7}^{{\left\{ \mathbb{H}, \mathbb{D} \right\}}}. \label{eq:AHD13}
\end{align}
This concludes the calculation of the \textit{complete} bracket $\{\mathbb{H},\mathbb{D}\}$. \\

As for $\{\mathbb{H},\mathbb{H}\}$, let us begin with the geometrical contribution,
\begin{subequations}
\begin{align}
     \biggl\{\mathbb{H}_\frak{g}[N_1], \mathbb{H}_\frak{g}[N_2]\biggr\}&=\bigl[ 1 + \alpha_3 \bigr] \bigl[ 1 + \alpha_5 \bigr] \, \mathbb{D}_\frak{g} \left[ \frac{\mathbf{N}}{\frak{p}}\biggl(\partial^a \ddN\biggr) \right] \\
     &+\intwK \mathbf{N} \, \ddN \delta K^b_b \, \mathcal{A}_1^{{\left\{ \mathbb{H}, \mathbb{H} \right\}}} \\
     &+ \intwK  \frac{\mathbf{N}}{2 \frak{p}} \, \ddN \delta E^b_b \, \mathcal{A}_2^{{\left\{ \mathbb{H}, \mathbb{H} \right\}}} \\
     &+ \intwK \mathbf{N} \,\bigl( \partial^b  \ddN \bigr) \bigl( \partial_a \delta^a_i \delta K^i_b \bigr)  \, \mathcal{A}_3^{{\left\{ \mathbb{H}, \mathbb{H} \right\}}} \\
     &+ \intwK \frac{\mathbf{N}}{\frak{p}} \,\bigl(  \partial^b \ddN \bigr) \bigl( \partial_a \delta^i_b \delta E^a_i \bigr)  \, \mathcal{A}_4^{{\left\{ \mathbb{H}, \mathbb{H} \right\}}},
\end{align}
\end{subequations}
with
\begin{align}
    \mathcal{A}_1^{{\left\{ \mathbb{H}, \mathbb{H} \right\}}} &= 2\bigl[ \tilde{g} + \alpha_1 \bigr] \biggl( g \ppc{g} - \tilde{g} - \alpha_6 \biggr)  - \biggl( g^2 + 4 \frak{p} g \ppp{g}\biggr) \poc{\tilde{g}+\alpha_1}  \notag \\
    &\qquad+ 4 \frak{p} g \ppc{g} \pop{\tilde{g}+\alpha_1} + \frac12 \bigl[ \tilde{g}^2 + \alpha_2 \bigr] \biggl( 2 + 3\alpha_5 - \alpha_4 \biggr) + \overline{\mathcal{A}}_{1}^{{\left\{ \mathbb{H}, \mathbb{H} \right\}}}, \label{eq:AHH1}\\
     \mathcal{A}_2^{{\left\{ \mathbb{H}, \mathbb{H} \right\}}} &= \bigl[ \tilde{g}^2 + \alpha_2 \bigr] \biggl( \tilde{g} + \alpha_6 - g \ppc{g} \biggr) - \frac12 \biggl( g^2 + 4 \frak{p} g \ppp{g} \biggr) \poc{\tilde{g}^2+\alpha_2} \notag\\
      &\qquad+ 2 \frak{p} g \ppc{g} \pop{\tilde{g}^2 + \alpha_2} + \bigl[ \tilde{g}+\alpha_1\bigr]\biggl( \tilde{g}^2 + 3 \alpha_8 - 2 \alpha_7\biggr) + \overline{\mathcal{A}}_{2}^{{\left\{ \mathbb{H}, \mathbb{H} \right\}}}, \label{eq:AHH2} \\
      \mathcal{A}_3^{{\left\{ \mathbb{H}, \mathbb{H} \right\}}} &= \bigl[ 1 + \alpha_3 \bigr] \bigl[ \alpha_5 - \alpha_4 \bigr], \label{eq:AHH3}\\
       \mathcal{A}_4^{{\left\{ \mathbb{H}, \mathbb{H} \right\}}} &= \bigl[ 1 + \alpha_3 \bigr] \biggl(\bigl[ \tilde{g} + \alpha_6 \bigr]  + \mathfrak{c} \bigl[ 1 + \alpha_5 \bigr]-g\ppc{g} \biggr) - \bigl[ \tilde{g} + \alpha_1 \bigr] \bigl[ 1 + \alpha_9 \bigr] + \notag \\
       &\qquad+ 2 \frak{p} g \ppc{g} \ppp{\alpha_3} + \biggl( \frac12 g^2 + 2 \frak{p} \ppp{g} \biggr) \ppc{\alpha_3} - \overline{\mathcal{A}}_{5}^{{\left\{ \mathbb{H}, \mathbb{H} \right\}}}. \label{eq:AHH4}
\end{align}
For the matter contribution, one obtains,
\begin{subequations}
\begin{align}
    \biggl\{\mathbb{H}_\frak{m}[N_1], \mathbb{H}_\frak{m}[N_2]\biggr\} &= \bigl[ 1+\beta_1 \bigr]\bigl[ 1 + \beta_{13} \bigr] \, \mathbb{D}_\frak{m}\biggl[\frac{\mathbf{N}}{\frak{p}}\biggl(\partial^a \ddN\biggr) \biggr] \\
    &+ \kappa \! \intnK \frac{\mathbf{N} \bm{\pi}}{2\frak{p}} \left(\frac{\bm{\pi}^2}{2 \frak{p}^3} - V[\bm{\phi}]\right) \ddN  \delta \pi \mathcal{A}_5^{{\left\{ \mathbb{H}, \mathbb{H} \right\}}} \\
    &+ \kappa \! \intnK \frac{\mathbf{N}}{2}  \, \partial_\phi \bigl( V[\bm{\phi}] \bigr) \left( \frac{\bm{\pi}^2}{2 \frak{p}} - \frak{p}^2 \, V[\bm{\phi}] \right) \ddN \, \delta \phi \mathcal{A}_6^{{\left\{ \mathbb{H}, \mathbb{H} \right\}}} \\
    &+ \kappa \! \intnK \frac{\mathbf{N} \bm{\pi}^4}{16 \frak{p}^5} \, \ddN \delta E^b_b \mathcal{A}_7^{{\left\{ \mathbb{H}, \mathbb{H} \right\}}} \\
    &+ \kappa \! \intnK \frac{\mathbf{N} \frak{p}}{4} V^2 \, \ddN \delta E^b_b \mathcal{A}_8^{{\left\{ \mathbb{H}, \mathbb{H} \right\}}} \\
    &+ \kappa \! \intnK \frac{\mathbf{N} \bm{\pi}^2}{8 \frak{p}^2}  V \, \ddN \delta E^b_b \mathcal{A}_9^{{\left\{ \mathbb{H}, \mathbb{H} \right\}}} \\
    &+ \intnK \mathbf{N} \, \bigl(\partial_\phi  V \bigr) \, \ddN \delta \pi \mathcal{A}_{10}^{{\left\{ \mathbb{H}, \mathbb{H} \right\}}} \\
    &+ \intnK \mathbf{N} \bm{\pi} \,  \bigl(\partial^2_\phi V \bigr) \, \ddN \delta \phi \mathcal{A}_{11}^{{\left\{ \mathbb{H}, \mathbb{H} \right\}}} \\
    &+ \intnK \frac{\mathbf{N} \bm{\pi}}{2 \frak{p}} \, \bigl(\partial_\phi V \bigr) \, \ddN \delta E^b_b \mathcal{A}_{12}^{{\left\{ \mathbb{H}, \mathbb{H} \right\}}},
\end{align}
\end{subequations}
where
\begin{align}
    \mathcal{A}_5^{{\left\{ \mathbb{H}, \mathbb{H} \right\}}} &= \partial_\frak{c}\beta_1 -  \overline{\mathcal{A}}_{20}^{{\left\{ \mathbb{H}, \mathbb{H} \right\}}}, \label{eq:AHH5}\\
    \mathcal{A}_6^{{\left\{ \mathbb{H}, \mathbb{H} \right\}}} &= \partial_\frak{c}\beta_3 - \overline{\mathcal{A}}_{25}^{{\left\{ \mathbb{H}, \mathbb{H} \right\}}} + \overline{\mathcal{A}}_{26}^{{\left\{ \mathbb{H}, \mathbb{H} \right\}}}, \label{eq:AHH6}\\
    \mathcal{A}_7^{{\left\{ \mathbb{H}, \mathbb{H} \right\}}} &= -\partial_\frak{c}\beta_2 +  \overline{\mathcal{A}}_{24}^{{\left\{ \mathbb{H}, \mathbb{H} \right\}}}, \label{eq:AHH7}\\
    \mathcal{A}_8^{{\left\{ \mathbb{H}, \mathbb{H} \right\}}} &= -\partial_\frak{c}\beta_4 + \overline{\mathcal{A}}_{27}^{{\left\{ \mathbb{H}, \mathbb{H} \right\}}}, \label{eq:AHH8}\\
    \mathcal{A}_9^{{\left\{ \mathbb{H}, \mathbb{H} \right\}}} &= \poc{\beta_2 + \beta_4}  +\overline{\mathcal{A}}_{23}^{{\left\{ \mathbb{H}, \mathbb{H} \right\}}}, \label{eq:AHH9} \\
    \mathcal{A}_{10}^{{\left\{ \mathbb{H}, \mathbb{H} \right\}}} &= \beta_1 - \beta_3 - \beta_5 - \beta_3 \beta_5, \label{eq:AHH10}\\
    \mathcal{A}_{11}^{{\left\{ \mathbb{H}, \mathbb{H} \right\}}} &= \beta_1 + \beta_7 + \beta_1 \beta_7 - \beta_3, \label{eq:AHH11}\\
    \mathcal{A}_{12}^{{\left\{ \mathbb{H}, \mathbb{H} \right\}}} &= \beta_1 + \beta_3 + \beta_6 + \beta_3 \beta_6 + \beta_8 + \beta_1 \beta_8 - \beta_2 - \beta_4 +  \overline{\mathcal{A}}_{19}^{{\left\{ \mathbb{H}, \mathbb{H} \right\}}}. \label{eq:AHH12}
\end{align}
For the mixed Poisson bracket involving the geometric and matter components of the Hamiltonian constraint, denoted by $\Delta \{\mathbb{H}_\frak{g}, \mathbb{H}_\frak{m}\}$, we get,
\begin{subequations}
\begin{align}
      \Delta \biggl\{\mathbb{H}_\frak{g}, \mathbb{H}_\frak{m}\biggr\} &= \intnK \frac{\mathbf{N} \bm{\pi}^2}{ 4 \frak{p}^2} \, \ddN \delta K^b_b \mathcal{A}_{13}^{{\left\{ \mathbb{H}, \mathbb{H} \right\}}} \\
     &+ \intnK \frac{\mathbf{N} \frak{p}}{2} V \, \ddN \delta K^b_b \mathcal{A}_{14}^{{\left\{ \mathbb{H}, \mathbb{H} \right\}}} \\
     &+ \intnK \frac{\mathbf{N}}{2} \biggl( \frac{\bm{\pi}^2}{2 \frak{p}^3} - V \biggr)\bigl( \partial^a  \ddN \bigr)\bigl( \partial_b  \delta^i_a \delta E^b_i \bigr) \mathcal{A}_{15}^{{\left\{ \mathbb{H}, \mathbb{H} \right\}}} \\
     &+ \intnK \frac{\mathbf{N} \bm{\pi}}{2 \frak{p}^2} \, \ddN \delta \pi \mathcal{A}_{16}^{{\left\{ \mathbb{H}, \mathbb{H} \right\}}} \\
     &+ \intnK \frac{\mathbf{N} \bm{\pi}^2}{8 \frak{p}^3} \ddN \delta E^b_b \mathcal{A}_{17}^{{\left\{ \mathbb{H}, \mathbb{H} \right\}}} \\
     &+ \intnK \frac{\mathbf{N}}{4} V \, \ddN \delta E^b_b \mathcal{A}_{18}^{{\left\{ \mathbb{H}, \mathbb{H} \right\}}} \\
     &+ \intnK \mathbf{N} \, \bigl( \partial_\phi V \bigr) \ddN \delta \phi \mathcal{A}_{19}^{{\left\{ \mathbb{H}, \mathbb{H} \right\}}},
\end{align}
\end{subequations}
where seven new anomalies appear,
\begin{align}
    \mathcal{A}_{13}^{{\left\{ \mathbb{H}, \mathbb{H} \right\}}} &=  \bigl[ 1 + \beta_2 \bigr]\bigl( 2 - \alpha_4 + 3 \alpha_5 \bigr) - 2 \poc{\tilde{g}+ \alpha_1} +  \overline{\mathcal{A}}_{12}^{{\left\{ \mathbb{H}, \mathbb{H} \right\}}}, \label{eq:AHH13}\\
    \mathcal{A}_{14}^{{\left\{ \mathbb{H}, \mathbb{H} \right\}}} &= \bigl[ 1 + \beta_4 \bigr]\bigl( -2 + \alpha_4 - 3 \alpha_5 \bigr) + 2 \poc{\tilde{g}+ \alpha_1} +  \overline{\mathcal{A}}_{11}^{{\left\{ \mathbb{H}, \mathbb{H} \right\}}} ,\label{eq:AHH14}\\
    \mathcal{A}_{15}^{{\left\{ \mathbb{H}, \mathbb{H} \right\}}} &= \ppc{\alpha_3} +  \overline{\mathcal{A}}_{13}^{{\left\{ \mathbb{H}, \mathbb{H} \right\}}} +  \overline{\mathcal{A}}_{14}^{{\left\{ \mathbb{H}, \mathbb{H} \right\}}},\label{eq:AHH15}\\
    \mathcal{A}_{16}^{{\left\{ \mathbb{H}, \mathbb{H} \right\}}} &= - 6\bigl[ \tilde{g}+ \alpha_1 \bigr] \bigl[ 1 + \beta_6 \bigr] + 2g \ppc{g} \bigl( 3 + 3 \beta_1 - 2 \frak{p} \ppp{\beta_1} \bigr) + g^2 \ppc{\beta_1} + 4 g \frak{p} \ppp{g} \ppc{\beta_1} +  \overline{\mathcal{A}}_{9}^{{\left\{ \mathbb{H}, \mathbb{H} \right\}}}, \label{eq:AHH16} \\
    \mathcal{A}_{17}^{{\left\{ \mathbb{H}, \mathbb{H} \right\}}} &= 12 \tilde{g} + 10 \alpha_1 + 2 \alpha_6 + 4 \bigl[\tilde{g}+\alpha_1\bigr] \beta_{10} + 2 \bigl[\tilde{g}+\alpha_6\bigr] \beta_2  + 6 \tilde{g} \beta_9 - 2 \tilde{g} \ppc{\tilde{g}} \notag\\ 
    &\qquad - 2 g \ppc{g} \bigl( 5 + 5 \beta_2 - 2 \frak{p} \ppp{\beta_2} \bigr)  - \ppc{\alpha_2} - g^2 \ppc{\beta_2} - 4 g \frak{p} \ppp{g} \ppc{\beta_2} +  \overline{\mathcal{A}}_{8}^{{\left\{ \mathbb{H}, \mathbb{H} \right\}}},\label{eq:AHH17}\\
    \mathcal{A}_{18}^{{\left\{ \mathbb{H}, \mathbb{H} \right\}}} &= 2 \alpha_1 - 2 \alpha_6 + 6 \bigl[\tilde{g} + \alpha_1 \bigr] \beta_{11} - 4 \bigl[\tilde{g}+\alpha_1\bigr] \beta_{12} - 2 \bigl[\tilde{g} + \alpha_6 \bigr]\beta_4 + 2 \tilde{g} \ppc{\tilde{g}} \notag \\
    &\qquad - 2 g \ppc{g} \bigl( 1 + \beta_4 + 2 \frak{p} \ppp{\beta_4} \bigr) + \ppc{\alpha_2} + g^2 \ppc{\beta_4} + 4 g \frak{p} \ppp{g} \ppc{\beta_4} +  \overline{\mathcal{A}}_{7}^{{\left\{ \mathbb{H}, \mathbb{H} \right\}}}, \label{eq:AHH18}\\
    \mathcal{A}_{19}^{{\left\{ \mathbb{H}, \mathbb{H} \right\}}} &= 3 \frak{p} \bigl[ \tilde{g} + \alpha_1 \bigr] + 3 \frak{p} \bigl[\tilde{g}+\alpha_1\bigr] \beta_8- g \ppc{g} \bigl( 3 \frak{p} \bigl[ 1 + \beta_3 \bigr] + 2 \frak{p}^2 \ppp{\beta_3} \bigr) + \frac{\frak{p}}{2} g^2 \ppc{\beta_3}\notag\\
    &\qquad  + 2 \frak{p}^2 g \ppp{g} \ppc{\beta_3} +  \overline{\mathcal{A}}_{10}^{{\left\{ \mathbb{H}, \mathbb{H} \right\}}}. \label{eq:AHH19}
\end{align}

Starting with the anomalies $\mathcal{A}_{8}^{\{\mathbb{H}, \mathbb{D}\}}$ (\ref{eq:AHD8}), $\mathcal{A}_{9}^{\{\mathbb{H}, \mathbb{D}\}}$ (\ref{eq:AHD9}), $\mathcal{A}_{10}^{\{\mathbb{H}, \mathbb{D}\}}$ (\ref{eq:AHD10}) we obtain, respectively,
\begin{equation}
    \beta_4 = 0, \quad \beta_{10} = 0, \quad \text{and} \quad \beta_{12}=0.
    \label{eq:b4b10b12}
\end{equation}

Equation \eqref{eq:b4b10b12} and $\mathcal{A}_{13}^{\{\mathbb{H}, \mathbb{D}\}}$ (\ref{eq:AHD13}) lead to,
\begin{equation}
    \beta_{11} = 0.
\end{equation}

Considering $\mathcal{A}_{3}^{\{\mathbb{H}, \mathbb{H}\}}$ (\ref{eq:AHH3}) adds,
\begin{equation}
    \alpha_5 = \alpha_4.
    \label{eq:a5eqa4}
\end{equation}

Thanks to Eq. \eqref{eq:pcb1234}, we automatically cancel  $\mathcal{A}_{5}^{\{\mathbb{H}, \mathbb{H}\}}$ (\ref{eq:AHH5}), $\mathcal{A}_{6}^{\{\mathbb{H}, \mathbb{H}\}}$ (\ref{eq:AHH6}), $\mathcal{A}_{7}^{\{\mathbb{H}, \mathbb{H}\}}$ (\ref{eq:AHH7}), $\mathcal{A}_{8}^{\{\mathbb{H}, \mathbb{H}\}}$ (\ref{eq:AHH8}), and $\mathcal{A}_{9}^{\{\mathbb{H}, \mathbb{H}\}}$ (\ref{eq:AHH9}). \\

Due to Eq. (\ref{eq:alpha3sol}), $\mathcal{A}_{15}^{\{\mathbb{H}, \mathbb{H}\}}$ (\ref{eq:AHH15}) vanishes. \\

Using Eq. \eqref{eq:pcb1234}, Eq. \eqref{eq:alpha3sol}, and Eq. \eqref{eq:a5eqa4}, one can easily obtain from considering $\mathcal{A}_{14}^{\{\mathbb{H}, \mathbb{H}\}}$ (\ref{eq:AHH14}) that
\begin{equation}
    \alpha_5 = \bigl(\pc \tilde{g}\bigr) - 1,
    \label{eq:a5sol}
\end{equation}
which, in turn, combined with $\mathcal{A}_{13}^{\{\mathbb{H}, \mathbb{H}\}}$ (\ref{eq:AHH13}), implies
\begin{equation}
    \beta_2 = 0,
    \label{eq:b2sol}
\end{equation}
as $\tilde{g}$ cannot be independent of the reduced curvature $\frak{c}$ by construction. Since $\beta_2=0$, one can derive from $\mathcal{A}_{7}^{\{\mathbb{H}, \mathbb{D}\}}$ (\ref{eq:AHD7}),
\begin{equation}
    \beta_1 = 0.
    \label{eq:b1sol}
\end{equation}
Equation \eqref{eq:b1sol} and $\mathcal{A}_{11}^{\{\mathbb{H}, \mathbb{H}\}}$ (\ref{eq:AHH11}) lead to,
\begin{equation}
    \beta_3 = \beta_7.
\end{equation}
Equation \eqref{eq:b1sol} and $\mathcal{A}_{6}^{\{\mathbb{H}, \mathbb{D}\}}$ (\ref{eq:AHD6}) lead to,
\begin{equation}
    \beta_5 = \beta_6.
\end{equation}
Equation \eqref{eq:b1sol} and $\mathcal{A}_{10}^{\{\mathbb{H}, \mathbb{H}\}}$ (\ref{eq:AHD10}) lead to,
\begin{equation}
    \beta_3 = -\frac{\beta_5}{1+\beta_5}.
\end{equation}
Equation \eqref{eq:b4b10b12}, Eq. \eqref{eq:b2sol}, Eq. \eqref{eq:b1sol}, and $\mathcal{A}_{6}^{\{\mathbb{H}, \mathbb{D}\}}$ (\ref{eq:AHH11}) imply:
\begin{equation}
    2 \beta_6 = \beta_9.
\end{equation}

As for $\beta_8$, using Eq. \eqref{eq:pcb1234} and $\mathcal{A}_{12}^{\{\mathbb{H}, \mathbb{D}\}}$ (\ref{eq:AHD12}), one obtains
\begin{equation}
    \beta_8 = -4 \frak{p}^2 \bigl( \pp \beta_3 \bigr),
\end{equation}
which, thanks to $\mathcal{A}_{12}^{\{\mathbb{H}, \mathbb{H}\}}$ (\ref{eq:AHH12}), leads to the condition
\begin{equation}
    \bigl( \pp \beta_3 \bigr) \biggl[ 2 \frak{p} - 4\frak{p^2} \biggr] = 0.
\end{equation}
As $\frak{p}$ can obviously not be constant in a dynamical model of the Universe, one is led to
\begin{equation}
     \bigl( \pp \beta_3 \bigr) = 0.
     \label{eq:ppb3}
\end{equation}
Combined with Eq. \eqref{eq:pcb1234}, this implies -- considering the classical limit -- that
\begin{equation}
    \beta_3 = 0,
\end{equation}
which, considering the relations with other matter counterterms, leads to,
\begin{equation}
    \beta_5 = 0, \quad \beta_6=0, \quad \beta_7=0, \quad \beta_8=0, \quad\text{and}\quad\beta_9=0.
\end{equation}
In addition, $\alpha_1$ can be easily determined using $\mathcal{A}_{16}^{\{\mathbb{H}, \mathbb{H}\}}$ (\ref{eq:AHH16}),
\begin{equation}
    \alpha_1 = g \bigl( \pc g \bigr) - \tilde{g}.
    \label{eq:a1sol}
\end{equation}
This is a very important restriction on $\alpha_1$ that has consequences on the generalized holonomy corrections that could be used to get a consistent theory. Thanks to Eq. \eqref{eq:pca1}, we know that
\begin{equation}
    \alpha_1 = \alpha_1(\frak{p}),
\end{equation}
which restricts the possible generalized corrections satisfying $\mathcal{A}_{16}^{\{\mathbb{H}, \mathbb{H}\}}=0$.\\ 

To go ahead, we now consider two specific important cases. First, let us assume that $\tilde{g}=\frak{c}$. The generalized holonomy correction $g$ is given by
\begin{equation}
    g\bigl(\frak{c}, \frak{p} \big) = \sqrt{\frak{c}^2 + 2 \frak{c} \alpha_1\bigl( \frak{p} \bigr) + f\bigl(\frak{p}\bigr)},
    \label{eq:main1}
\end{equation}
where $f(\frak{p})$ is a function of the reduced densitized triad $\frak{p}$ \textit{only}, which must vanish in the classical limit. 

The second significant case is $\tilde{g}=g$. In this case, the solution for $g$ is given by
\begin{equation}
    g\bigl(\frak{c}, \frak{p} \big) = \alpha_1\bigl( \frak{p} \bigr) \biggl[ 1 + W\biggl(- \alpha_1^{-1} \exp \biggl[ - \frac{\frak{c} + \alpha_1 + f\bigl(\frak{p}\bigr)}{\alpha_1} \biggr] \biggr)\biggr],
    \label{eq:main2}
\end{equation}
where $W(x)$ is the Lambert $W$ function. It should be noted that $g$ cannot be expressed in terms of the usual holonomy correction. In other words, we can already conclude that the usual substitution cannot be implemented consistently while expanding the functional dependence of the counterterms. This is a significant and, maybe, surprising conclusion. 

At this stage, Eqs. (\ref{eq:main1}) and (\ref{eq:main2}) are the main results of this work. They are different from all previously known restrictions.\\

These two scenarios require a careful treatment when anomalies exhibit an interplay between $g$ and $\tilde{g}$. However, at this stage, we can establish some general conditions to be fulfilled by the counterterms.\\

Using Eq.~\eqref{eq:a1sol} and  $\mathcal{A}_{1}^{\{\mathbb{H}, \mathbb{D}\}}$ (\ref{eq:AHD1}), we obtain,
\begin{equation}
    \alpha_2 = 3 g^2 - \tilde{g}^2 - 2 \frak{c} g \bigl( \pc g \bigr).
\end{equation}
Using the solution for $\alpha_5$ given by Eq. \eqref{eq:a5sol}, from $\mathcal{A}_{5}^{\{\mathbb{H}, \mathbb{D}\}}$ (\ref{eq:AHD5}) we get,
\begin{equation}
    \alpha_6 = 2 g \bigl( \pc g \bigr) - \tilde{g} - \frak{c} \bigl( \pc \tilde{g} \bigr),
    \label{eq:a6sol}
\end{equation}
which, automatically cancels $\mathcal{A}_{3}^{\{\mathbb{H}, \mathbb{D}\}}$ (\ref{eq:AHD3}) as well.\\

Via Eq. \eqref{eq:a6sol} and $\mathcal{A}_{4}^{\{\mathbb{H}, \mathbb{D}\}}$ (\ref{eq:AHD4}), we obtain 
\begin{equation}
    \alpha_7 = 4 g \biggl[ \frak{c} \bigl( \pc g \bigr) - \frak{p} \bigl( \pp g \bigr) \biggr] - 2 \frak{c}^2 \bigl( \pc \tilde{g} \bigr) - g^2 - \tilde{g}^2.
    \label{eq:a7sol}
\end{equation}
From the cancelation of $\mathcal{A}_{2}^{\{\mathbb{H}, \mathbb{D}\}}$ (\ref{eq:AHD2}), we then have,
\begin{equation}
    \alpha_8 = \alpha_7 + 4 \frak{p} \bigl( \pp \alpha_2 \bigr),
\end{equation}
while $\mathcal{A}_{4}^{\{\mathbb{H}, \mathbb{H}\}}$ (\ref{eq:AHH4}) implies
\begin{equation}
    \alpha_9 = 0.
    \label{eq:a9sol}
\end{equation}
This is interesting as this fixes the ambiguities on the counterterms $\alpha_3$ and $\alpha_9$ present in the flat-FLRW case with counterterms depending only on the reduced variables $\bigl(\frak{c}, \frak{p}\bigr)$. \\

The last counterterms to be fixed is $\beta_{13}$. It can be fixed ensuring that the bracket $\left\{\mathbb{H}, \mathbb{H}\right\}$ remains first class,
\begin{equation}
    \beta_{13} = \bigl(\pc \tilde{g} \bigr) - 1.
\end{equation}
This makes a substantial difference with the case where counterterms are depending on $\bigl( \frak{c}, \frak{p} \bigr)$ only. 
It is obviously possible to have a nondeformed algebra of constraints. 
However, in the case $\tilde{g}=g$, the algebra of constraints is deformed but, now, by the first derivative of the holonomy correction, unlike the usual case where the second derivative (which is linked with the energy density of the Universe) appears. This could potentially have consequences on the Mukanov-Sasaki equation, and thus on the dynamics on cosmological perturbations. \\

Still, some anomalies remain to be canceled. First, thanks to the condition $\beta_3=0$, and using Eq. \eqref{eq:a1sol}, we automatically get rid of $\mathcal{A}_{19}^{\{\mathbb{H}, \mathbb{H}\}}$ (\ref{eq:AHH19}). \\

From  $\mathcal{A}_{17}^{\{\mathbb{H}, \mathbb{H}\}}$ (\ref{eq:AHH17}) and  $\mathcal{A}_{18}^{\{\mathbb{H}, \mathbb{H}\}}$ (\ref{eq:AHH18}), we are led to the following restriction:
\begin{equation}
    12 \tilde{g} - 10 g \bigl( \pc g \bigr) + 12 \alpha_1 = 0,
\end{equation}
which is satisfied given Eq. \eqref{eq:a1sol}. \\

At this point, only three anomalies remain: $\mathcal{A}_{1}^{\{\mathbb{H}, \mathbb{H}\}}$ (\ref{eq:AHH1}), $\mathcal{A}_{2}^{\{\mathbb{H}, \mathbb{H}\}}$ (\ref{eq:AHH2}), and $\overline{\mathcal{A}}_{3}^{\{\mathbb{H}, \mathbb{H}\}}$  (\ref{eq:AHH3}). The previously derived counterterms respectively impose the following restrictions\footnote{For simplicity, $\mathcal{A}_{1}^{\{\mathbb{H}, \mathbb{H}\}}$ (\ref{eq:AHH1}), $\mathcal{A}_{2}^{\{\mathbb{H}, \mathbb{H}\}}$ (\ref{eq:AHH2}) are given modulo $\overline{\mathcal{A}}_{3}^{\{\mathbb{H}, \mathbb{H}\}}$.},
\begin{equation}
    3 \bigl( \pc g\bigr)^2 - 3 \bigl( \pc \tilde{g} \bigr) - 4 \mathfrak{p} \bigl( \pc g \bigr) \bigl( \partial^2_{\mathfrak{c}, \mathfrak{p}} g \bigr) + g \bigl( \partial^2_\mathfrak{c} g \bigr) + 4 \mathfrak{p} \bigl( \pp g \bigr) \bigl( \partial^2_\mathfrak{c} g \bigr) = 0,
    \label{eq:simpleA1HH}
\end{equation}
\begin{subequations}
\begin{align}
    g^2 \bigl( \pc g \bigr) + \tilde{g}^2 \bigl( \pc g \bigr) &+ 4 \mathfrak{p} g \bigl( \pp g \bigr) \bigl( \pc g \bigr) - 7 \mathfrak{c} g \bigl( \pc g \bigr)^2 + 3 \mathfrak{c} g \bigl( \pc \tilde{g} \bigr) + 2 \mathfrak{c}^2 \bigl( \pc g \bigr) \bigl( \pc \tilde{g} \bigr) \\ 
    &+ 4 \mathfrak{c} \mathfrak{p} g \bigl( \pc g \bigr) \bigl( \partial^2_{\mathfrak{c}, \mathfrak{p}} g \bigr) - \mathfrak{c} g^2 \bigl( \partial^2_\mathfrak{c} g \bigr) - 4 \mathfrak{c} \mathfrak{p} g \bigl( \pp g \bigr) \bigl( \partial^2_\mathfrak{c} g \bigr) = 0,
\end{align} 
\label{eq:simpleA2HH}
\end{subequations}
and
\begin{equation}
    2 \mathfrak{p} \bigl(\pp\tilde{g}\bigr) - 2 g \bigl( \pc g \bigr) - 2 \mathfrak{p} \bigl( \pp g\bigr) \bigl( \pc g \bigr) + \mathfrak{c} \bigl( \pc g \bigr)^2 + \tilde{g} \bigl( \pc \tilde{g} \bigr) - 2 \mathfrak{p} g \bigl( \partial^2_{\mathfrak{c},\mathfrak{p}} g \bigr) + \mathfrak{c} g \bigl( \partial^2_\mathfrak{c} g \bigr) = 0.
    \label{eq:simpleA3BHH}
\end{equation}

This set of equations constitutes a novel constraint on the form of the holonomy corrections $g$ and $\tilde{g}$, arising from the extended functional dependence of the counterterms considered in this work. Given those equations, it is clear that assessing whether a first-class constraint algebra can be achieved or not is a highly nontrivial task. For any choice of holonomy corrections $g$ and $\tilde{g}$, the compliance with the preceding set of differential equations has to be carefully examined.\\


As an illustrative example, let us consider the simplest case, namely, $\tilde{g}=\mathfrak{c}$. The use of Eq. (\ref{eq:simpleA3BHH}) yields the following restriction:

\begin{equation}
    \alpha_1 + \mathfrak{p} \biggl( \pp \alpha_1 \biggr) = 0,
\end{equation}
implying
\begin{equation}
    \alpha_1\bigl[\mathfrak{p}\bigr] = K \mathfrak{p}^{-1},
\end{equation}
with $K$ a constant. Moreover, taking advantage of this explicit expression of $\alpha_1$, Eq. (\ref{eq:simpleA1HH}) leads to
\begin{equation}
    f[\mathfrak{p}] - \mathfrak{p} \biggl( \pp f[\mathfrak{p}] \biggr) = - \frac64 \alpha_1^2\bigl[ \mathfrak{p} \bigr].
\end{equation}
We can therefore conclude that
\begin{equation}
    f \bigl[ \mathfrak{p} \bigr] = \frac12 \alpha_1^2\bigl[ \mathfrak{p} \bigr] + \tilde{K} \mathfrak{p}.
\end{equation}
Based on the consistency requirement, $g\rightarrow\mathfrak{c}$ in the classical limit, we conclude that $\tilde{K}=0$, which, finally, taking into account Eq. (\ref{eq:simpleA2HH}), leads to the final condition,
\begin{equation}
    2  \bigl( \mathfrak{c} \mathfrak{p} \bigr)^2  K + 3 \mathfrak{cp} K^2 + K^3 = 0,
\end{equation}
meaning that $K=0$. Consequently, to achieve a first-class constraint algebra while implementing genearalized holonomy corrections at the background level only, one has to take $g=\mathfrak{c}$, which is equivalent to having no corrections.\\

\section{Conclusion}

In this work, we have considered an extended functional dependence of the counterterms within the deformed algebra approach to loop quantum cosmology. Specifically, we have gone beyond the standard framework where  counterterms depend solely on the background variables -- {\it i.e.} $\alpha_i(\mathfrak{c},\mathfrak{p})$ -- and, instead, have allowed them to depend on the full set of geometric phase space variables -- $\alpha_i(K^i_a, E^a_i)$. This motivation arises from the deformed algebra approach’s objective to characterize the behavior of cosmological perturbations within the context of loop quantum cosmology. The counterterms involved in the calculation process cannot be reasonably expected to depend exclusively on background variables.

This generalization significantly alters the conventional conclusions. In particular, we have shown that the usual replacement $\mathfrak{c} \rightarrow \sin(\delta\mathfrak{c}) / \delta$ no longer fulfills the consistency conditions. In other words, the standard correction cannot be consistently implemented within this framework, where the functional dependency of the counterterms has been extended.\\

This work also provides a novel set of constraints on the allowed forms of holonomy corrections to ensure a first-class constraint algebra. Consequently, determining \textit{a priori} whether a given holonomy correction yields a consistent theory becomes highly nontrivial.\\


In future works, beyond exploring the potential phenomenological implications of this approach, it would be important to examine other relevant functional dependencies of the counterterms, such as a possible dependence upon the matter sector.


\bibliographystyle{JHEP}
\bibliography{biblio.bib}

\end{document}